\documentclass[journal,compsoc]{IEEEtran}
\usepackage{cite}
\usepackage{graphicx}
\graphicspath{{./fig/}}
\usepackage{subfigure}
\usepackage{amsfonts}
\usepackage{multirow}
\usepackage{flushend}
\usepackage{soul}
\usepackage{diagbox}
\usepackage{booktabs}
\newcommand{\ra}[1]{\renewcommand{\arraystretch}{#1}}
\usepackage{xcolor}

\begin{document}

\title{Automatic Rumor Detection on Microblogs: A Survey }
%
%
% author names and IEEE memberships
% note positions of commas and nonbreaking spaces ( ~ ) LaTeX will not break
% a structure at a ~ so this keeps an author's name from being broken across
% two lines.
% use \thanks{} to gain access to the first footnote area
% a separate \thanks must be used for each paragraph as LaTeX2e's \thanks
% was not built to handle multiple paragraphs
%\author{Zhiwei~Jin,~Juan~Cao,~Han Guo, and Yongdong~Zhang,~\IEEEmembership{Senior Member,~IEEE,}}
\author{Juan~Cao,~Junbo~Guo,~Xirong~Li,~Zhiwei~Jin,~Han~Guo, and Jintao~Li}

\maketitle

\IEEEpubid{This work has been submitted to the IEEE for possible publication. Copyright may be\newline transferred without notice, after which this version may no longer be accessible.}

\begin{abstract}
The ever-increasing amount of multimedia content on modern social media platforms are valuable in many applications. While the openness and convenience features of social media also foster many rumors online. Without verification, these rumors would reach thousands of users immediately and cause serious damages. Many efforts have been taken to defeat online rumors automatically by mining the rich content provided on the open network with machine learning techniques. Most rumor detection methods can be categorized in three paradigms: the hand-crafted features based classification approaches, the propagation-based approaches and the neural networks approaches. In this survey, we introduce a formal definition of rumor in comparison with other definitions used in literatures. We summary the studies of automatic rumor detection so far and present details in three paradigms of rumor detection. We also give an introduction on existing datasets for rumor detection which would benefit following researches in this area. We give our suggestions for future rumors detection on microblogs as a conclusion.

\end{abstract}


\section{Introduction}
The explosive development of contemporary social media platforms has witnessed their key role for reporting and propagating news in the modern society. According to a recent study from Pew Research, 62 percent of people receive news from social media, with 18 percent doing so very often \cite{Gottfried2016}. Social media users not only read news but also propagate and even produce immediate news on the social network. With millions of people serving as ``news sensors", news on social media is valuable for opinion mining and decision making. However, the convenience of publishing news also fosters the emergence of various rumors. According to a report in China~\cite{zhao2015enquiring}, over one third of trending events on microblogs contain fake information.

The widespread of rumors can pose a threat to the internet credibility and cause serious consequences in real life\cite{friggeri2014rumor}. Followings are some examples showing rumors that cause damages to political events, economy and social stability.

\begin{itemize}
  \item
  During the 2016 U.S. presidential election, candidates and their supporters were actively involved on Facebook and Twitter to do campaigns and express their opinions. However, as many as 529 different rumor stories pertaining to presidential candidates Donald Trump and Hillary Clinton were spreading on social media during the election \cite{jin2017Rumor}. These rumors reached millions of voters via social network promptly and potentially influenced the election.
  \item
  On April 23rd 2013, the official Twitter account of Associated Press was hacked to sent out a tweet claiming two explosions happened in the White House and the president got injured. Even though this rumor was quickly debunked, it still spread to millions of users and caused severe social panic, resulting in a dramatic crash of the stock market immediately \cite{Domm2013}.
  \item
  In March 2014, promptly after the emergency event ``\emph{Malaysia Airlines Flight MH370 Lost Contact}'', 92 different rumor stories were spread widely on Sina Weibo, the primary microblog service in China \cite{jin2014news}. These rumors blocked people from knowing the real situation of the event and damaged the feelings of family members related to the missing passengers on the plane.
\end{itemize}

\begin{figure}[!tb]
\centering
\subfigure[Fackebook]{

\includegraphics[width=\columnwidth]{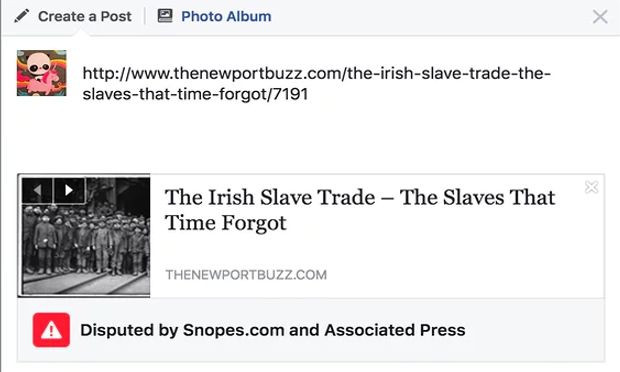}
\label{fig_1a}
}
\subfigure[Twitter]{
\includegraphics[width=\columnwidth]{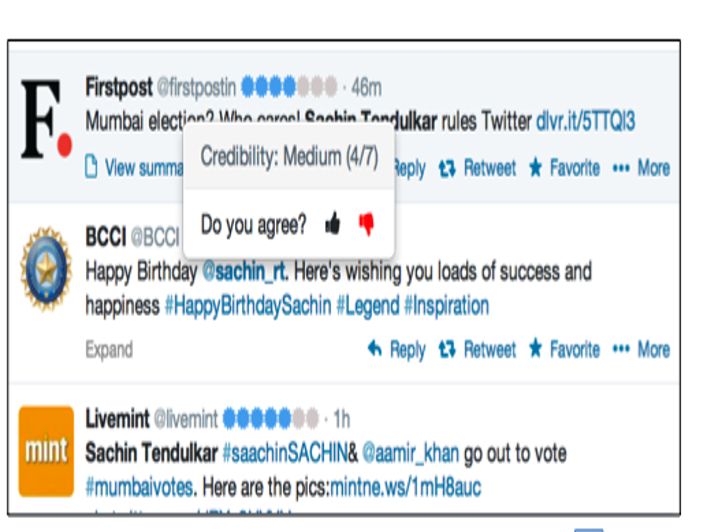}
\label{fig_1b}
}
\subfigure[Weibo]{

\includegraphics[width=\columnwidth]{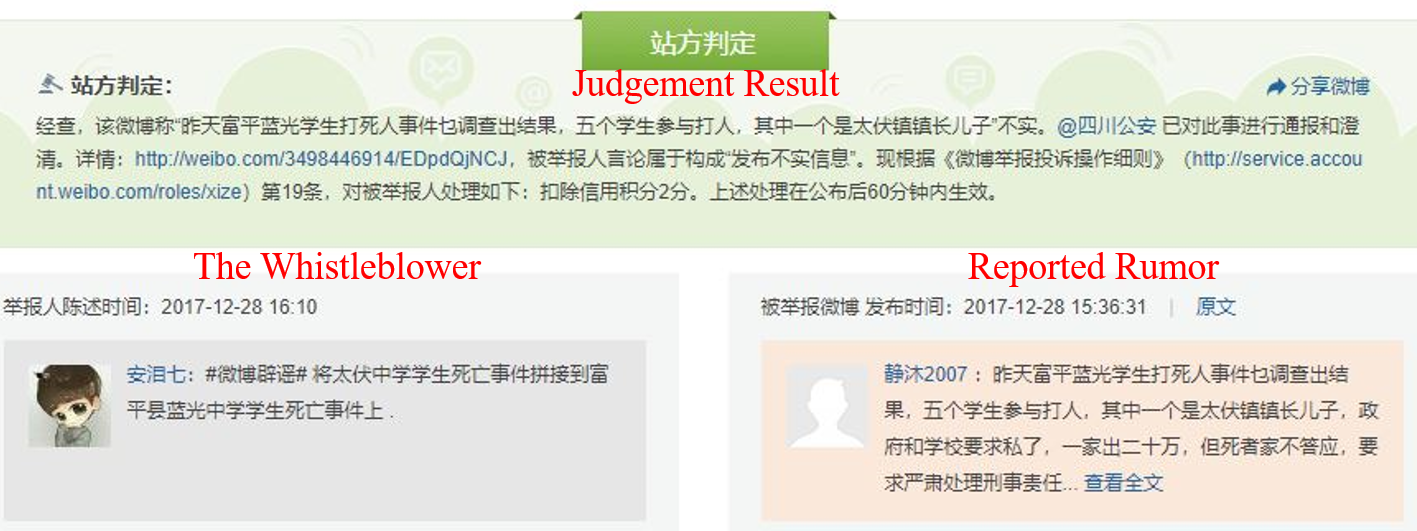}
\label{fig_1c}
}
\caption{An illustration of rumor combating strategies in three major social media platforms: (a) Facebook, (b) Twitter, and (c) Weibo.}
\label{fig_1}
\end{figure}

Rumors on social media have become a serious concern recently, especially when people are aware of their capabilities to influence the society. Commercial giants, government authorities and researchers are all taking great efforts in defeating the negative impacts of rumors.
%Under the large amount of the public criticism that social media platforms are the breeding grounds for rumors and fake news, many popular social media platforms are taking actions to combat rumors. 
Figure \ref{fig_1} showcases anti-rumor designs in three major platforms including Facebook, Twitter and Weibo. Facebook labels fake news stories with the help of users and outside fact checkers (Figure \ref{fig_1a}). Users alert the platform to possible rumors, which will be sent by the platform to fact-checking organizations such as AP, FactCheck.org and Snopes.com for verification. Verified fake story will be publicly flagged as disputed by a 3rd party fact-checkers whenever it appears on the social network. Users will get another warning if they insist on sharing verified fake stories \cite{Facebook2016}. Twitter employs a semi-automatic strategy combining automatic evaluation and crowd-sourcing annotation to flag possible fake tweets (Figure \ref{fig_1b}). Each tweet is assigned with a credibility rating automatically generated by an algorithm \cite{Gupta2014TweetCred}. Users are allowed to give their feedback if they disagree with the rating. On Weibo, users are encouraged to report possible fake tweets. These tweets are passed to a commit composed of elite users to scrutinize and judge them (Figure \ref{fig_1c}). Verified fake tweets will be labeled. The above methods depend on social media users or human experts to detect rumors. Moreover, these methods mostly focus on defeating rumors during their propagation process on social media, rather than detecting emerging rumors at an early stage.

\begin{figure*}[!tb]
\centering
\includegraphics[width=2\columnwidth]{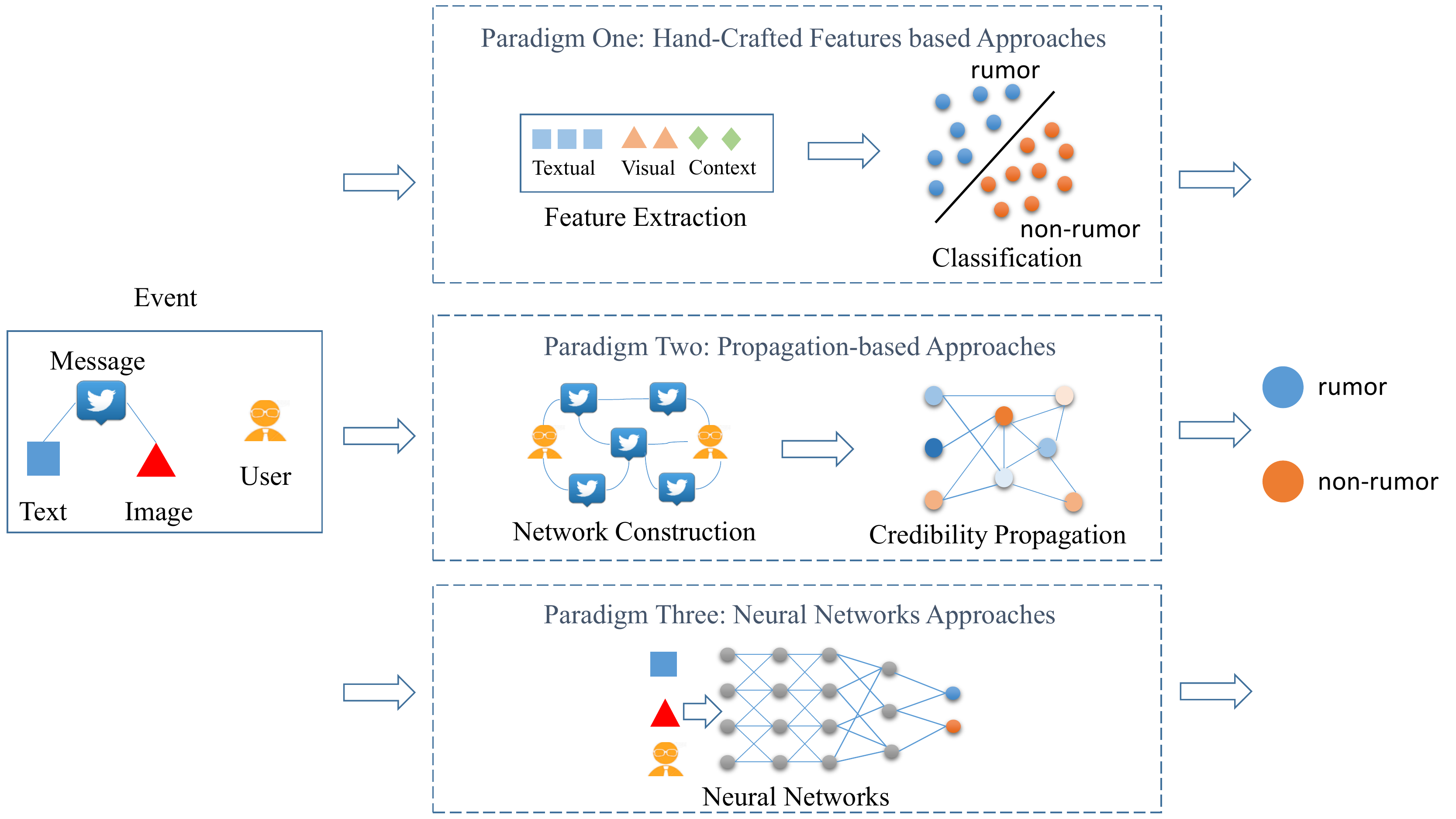}
\caption{Paradigms for automatic rumor detection.}

\label{fig_paradigms}
\end{figure*}

Although manual rumor annotation of large-scale social media data becomes feasible, manual labeling encounters the following challenges:
\begin{enumerate}
  \item Intensive labor and cost of time \cite{ma2016detecting}. Even with a good motivation, users may be frustrated by the intensive labor and cost of naive manual interactions to label rumors. Heavy labor and cost of time are obstacles for manually labeling of large-scale online data.
  \item Label quality. Unlike common human interactive annotations such as tagging pictures on image sharing websites \cite{Wang2012}, identifying a fresh rumor involves much more domain knowledge and elaborative examination. It is nontrivial for an journalism expert to fact-check a news story, let alone nonprofessional users. The low-quality labels will directly jeopardize the identification of rumors and fail to meet the qualification of accurate rumor detection.
\end{enumerate}

Due to the above challenges, most existing applications aim at simultaneously exploring humans and computers for rumor detection. In comparison with pure manual labeling, they are able to provide better results with high efficiency because labor cost is reduced, as users may not need to label all the data. 

Automatic rumor detection on social media faces many challenges, as the social multimedia data is numerous and ever-increasing. What's more, the non-structural, incomplete and noisy nature of online data makes it very challenging to process and understand. We summarize the challenges as follows.
\begin{enumerate}
  \item Semantics understanding. Most rumors are deliberately fabricated to misleading the public. Fake information is mixed and disguised in a rumor story. It is very difficult for machine to completely understand such semantics.
  \item Huge variations. Rumors can cover all kinds of topics and take various language styles. Features effective to distinguish certain type of rumors may work poorly on other types. Algorithms trained on limited labeled data would probably fail on unseen new rumors.
  \item Multimodal data. Rumors on social media often take the form of images or videos, apart from pure texts. Although information in different modalities can provide clues for rumor detection, how to extract prominent features from each modality and effectively fuse them together is challenging.
  \item Heterogeneous propagation structure. During the diffusion of rumors on the social network, users can discuss and make comments. The users' behaviors help to construct an underlying social context for rumors. Characterizing these behaviors on the propagation network is crucial to identify rumors online.
\end{enumerate}

Facing these challenges, good efforts have been made to attack the automatic rumor detection problem from different angles, as visualized in Fig. \ref{fig_paradigms}. Most approaches formulate the automatic rumor detection as a two-class classification task. We divide these approaches into three categories: handcrafted features based approaches, propagation-based approaches and neural networks approaches.
\begin{enumerate}
\item Category I: handcrafted features based approaches. The traditional methods apply hand-crafted features to describe the distribution of rumors in high dimensional space and the separating hyperplane will be learned by classifiers\cite{castillo2011information,yang2012automatic,kwon2013prominent,sun2013detecting,wu2015false,jin2015}. These studies extract features from the textual and visual content of rumors. To avoid the variations in content semantics, some social context features are also proposed to capture the characteristics of rumors, so the feature engineering becomes a fatal step before classification. However, as some rumors lack of some crucial features, these methods often lead to unstable and unreliable results.
\item Category II: propagation based approaches. To utilize the heterogeneous structure of social network, Graph-based optimization methods link messages and users into a whole network and evaluate their credibilities as a whole \cite{gupta2012evaluating}\cite{jin2014news}\cite{jin2016}. The category of a unclassified text can be obtained when the network is convergent. However, it is obviously to see that these work ignore the textual information represented by rumors.
\item Category III: neural network approaches. More recently, deep neural networks are proposed to learn and fuse multi-modal features automatically for rumor detection \cite{ma2016detecting,chen2017call,ruchansky2017csi,jin2017multimodal,yuconvolutional,nguyen2017early}. By modeling text data as time-series, either recurrent neural network or convolutional neural network can learn the latent textual representation and improve the accuracy of classification. Compared to the work that leverage traditional classifiers, these work can significantly improve the performance.
\end{enumerate}

Several survey papers exist. Zubiaga \emph{et al.} propose an overview of how to develop a rumor detection system, which consists of four steps: rumor detection, rumor tracking, rumor stance classification, and rumor veracity classification \cite{zubiaga2017detection}. As their work pays less attention on analyzing the feature engineering step and neural network based algorithms, which can not summarize and compare different methods from the perspective of algorithms. Boididou \emph{et al.} empirically compare three methods on a dataset collected form Twitter \cite{boididou2017verifying}, without considering the development of rumor detection on Weibo, which ignore some prominent researches\cite{jin2014news,yang2012automatic,jin2015}. To sum up, the previous lacks of comprehensive analysis for detection algorithms and ignores the development of neural network on rumor detection, which motivates our work. Most rumor detection methods can be modeled as a two-step work, consisting of feature selection and binary classification. Depending on the difference in selecting features and machine learning algorithms among the present researches, we aim to provide a comprehensive introduction for existing rumor detection algorithms. Specifically, we categorize the methods described in over 50 papers into three paradigms: hand-crafted feature based methods, propagation based methods and neural network based methods. We analyze the advantages and weakness of each paradigm as well. Such an analysis helps to point out the application context for each algorithm.

The rest of the article is organized as follows. In Section 2, we introduce the formal detection of rumor detection. We then introduce the work on the classification-based approaches for rumor detection in Section 3. We give our a detailed introduction of categories of prominent features extracted for rumor detection models followed by classification methods integrating these features for the rumor detection task in this section. Two types of state-of-the-art rumor detection models, propagation-based models and deep neural networks models are then introduced in section 4 and 5, respectively. In Section 6, we discuss the dataset status for the research of rumor detection. Finally, we conclude the paper and discuss future research directions in Section 7.
%-----------------------------------------------
\section{Definitions} \label{sec:def}
%-----------------------------------------------

The definition of a rumor varies over publications. In \cite{difonzo2007rumor}, for instance, DiFonzo \emph{et al.} define rumor as a story or statement in general circulation without confirmation or certainty to facts. While in \cite{allport1947psychology}, a story or a statement whose truth value id unverified or deliberately false is considered as rumor. The lack of consistency in this fundamental concept makes a head-to-head comparison between existing methods difficult. In this section, we first introduce three typical rumor definitions used in the literature. We then give a formal definition of the rumor detection problem, which allows us to interpret existing methods in a unified framework.

We give an objective definition for rumors discussed in this survey as follows:

\textbf{Definition: Objective Rumor} This definition of rumor is strictly equivalent to verified fake information. Once a statement is confirmed to have false or fabricated content by authoritative sources, it is labeled as a rumor. This kind of rumor is also referred as ``false rumor'' \cite{wu2015false} or ``fake news'' \cite{jin2014news}\cite{jin2016}\cite{ruchansky2017csi}. 
%in literatures.

\begin{figure} [tb!]
\centering
\includegraphics[width=\columnwidth]{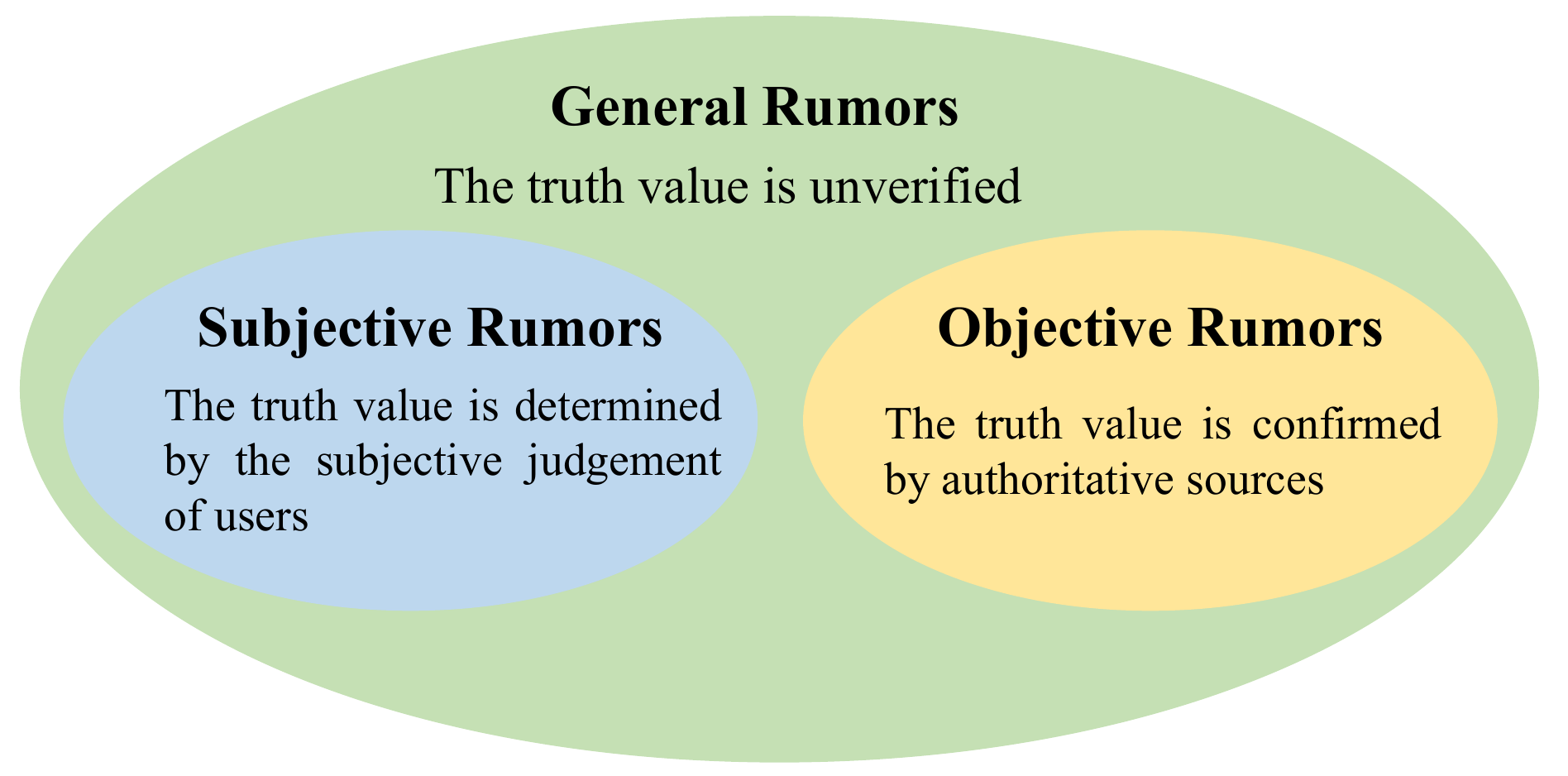}
\caption{Three definitions of rumors and their relationships}
\label{fig_definitions}
\end{figure}

Besides objective rumors, there also exist alternative definitions such as general rumors and subjective rumors, see Fig. \ref{fig_definitions}. A traditional definition of rumor is derived from the social psychology \cite{allport1947psychology}. That is, a rumor is an unconfirmed statement that widely spread and the truth value of which is unverified or deliberately false. Under this definition, the general rumor is deemed as a piece of information, the veracity of which has not been verified. Unverified rumors refer to information without official confirmation from authoritative or credible sources, e.g. authoritative news agencies or witnesses on site. Another definition for rumor is from the subjective judgment of users \cite{castillo2011information,gupta2012evaluating,kwon2013prominent}. An analogy of this definition is that the sentiment polarity of a statement is often defined based on people's subjective feelings.

Different definitions are better adopted in different scenarios. General rumors are often used in monitoring public opinions towards controversial stories on social media \cite{alexander2014social}. General rumors may be gossips of celebrities or a campaign strategy to slander political opponents~\cite{jin2017Rumor}, which attract people to spread the story rather than finding out its veracity. Some research~\cite{zhao2015enquiring} also suggests to use this definition to filter out unrelated posts and then classify the veracity of remained general rumors. Subjective rumors can be used to understand users' behavior. Based on this definition, Morris \emph{et al.} \cite{morris2012tweeting} study what kinds of messages are more likely to get the trust of the public on Twitter. Rumors with objective confirmation is widely used for automatic rumor detection \cite{yang2012automatic,jin2017multimodal,ma2016detecting,jin2017Novel}, methods for detecting verified rumors aim to detect false information at the early stage and thus prevent the hazard of rumors on social media.

Given our focus on detecting the veracity of a statement, we adopt the objective definition of rumors throughout the paper.
A news story $e$ is defined as a set of $n$ pieces of related messages $M=\{m_1, m_2, ... ,m_n\}$. For each messages $m_i$, it is comprised of a tuple of attributes representing the text, images, videos and other content it contains. Each message $m_i$ is also associated with a user $u_i$ who posted it. The user $u_i$ is represented with a set of attributes, including name, age, register time and avatar images, \emph{etc}. The rumor detection task is then defined as follows:

\textbf{Definition: Rumor Detection}: Given a news story $e$ with its message set $M$ and user set $U$, the rumor detection task aim to determine whether this story can be confirmed as true or false, i.e to learn a prediction function $\mathcal{F}\left(e\right) \rightarrow \{0, 1\}$ satisfying:

\begin{equation}
\mathcal{F}\left(e\right)
= \left\{ \begin{array}{l}
1,  if \ e \ is \ confirmed \ as\ false   \\
0,  otherwise
\end{array} \right.
\end{equation}

The definition formulates the rumor detection as a veracity classification task aiming to determine whether a given story on social media is confirmed as true or false. It reveals two main challenges in rumor detection: 1) the understanding of contents and social context associating with the story and 2) the constructing of effective detection algorithms. As illustrated in Figure \ref{fig_paradigms}, three paradigms are proposed focusing on extracting effective features representing rumor content or robust prediction algorithms. We summary these methods in the following three sections.
\section{Hand-crafted Features Based Approaches}

The basic definition of rumor detection is a binary classification problem. Most literatures on this task follows the commonly used learning paradigms of supervised classification from the machine learning area: 1) extracting features representing samples from both classes; 2) training an appropriate classifier with provided samples; 3) testing and evaluating on unseen data. The first key step of these procedures is how to extract prominent features. Moreover, contents are not independent and from different modalities. The classification method should be capable of integrating and interpreting these features for robust rumor detection. In this section, we first give a review of all kinds of features used in the rumor detection literatures. We then introduce typical classifiers applied for rumor detection. % literatures.

%\xr{Here comes question. You give a review of ALL kinds of features. Why not introducing ALL kinds of classifiers?}

\begin{figure} [tb!]
\centering
\includegraphics[width=\columnwidth]{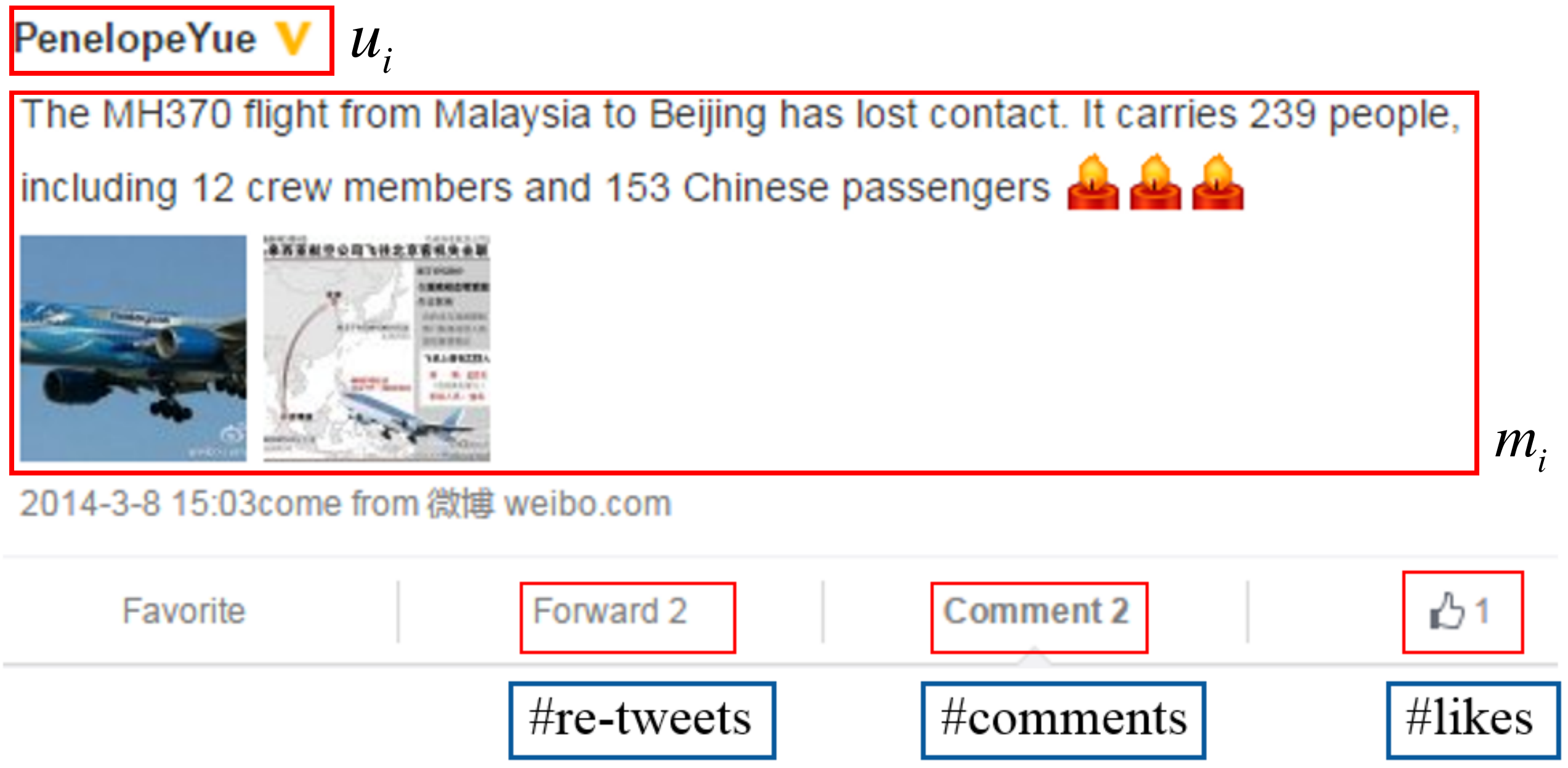}
\caption{An example message on Weibo with some of its key features and contents marked.}
\label{fig_example}
\end{figure}

\begin{figure}
\centering
\includegraphics[width=\columnwidth]{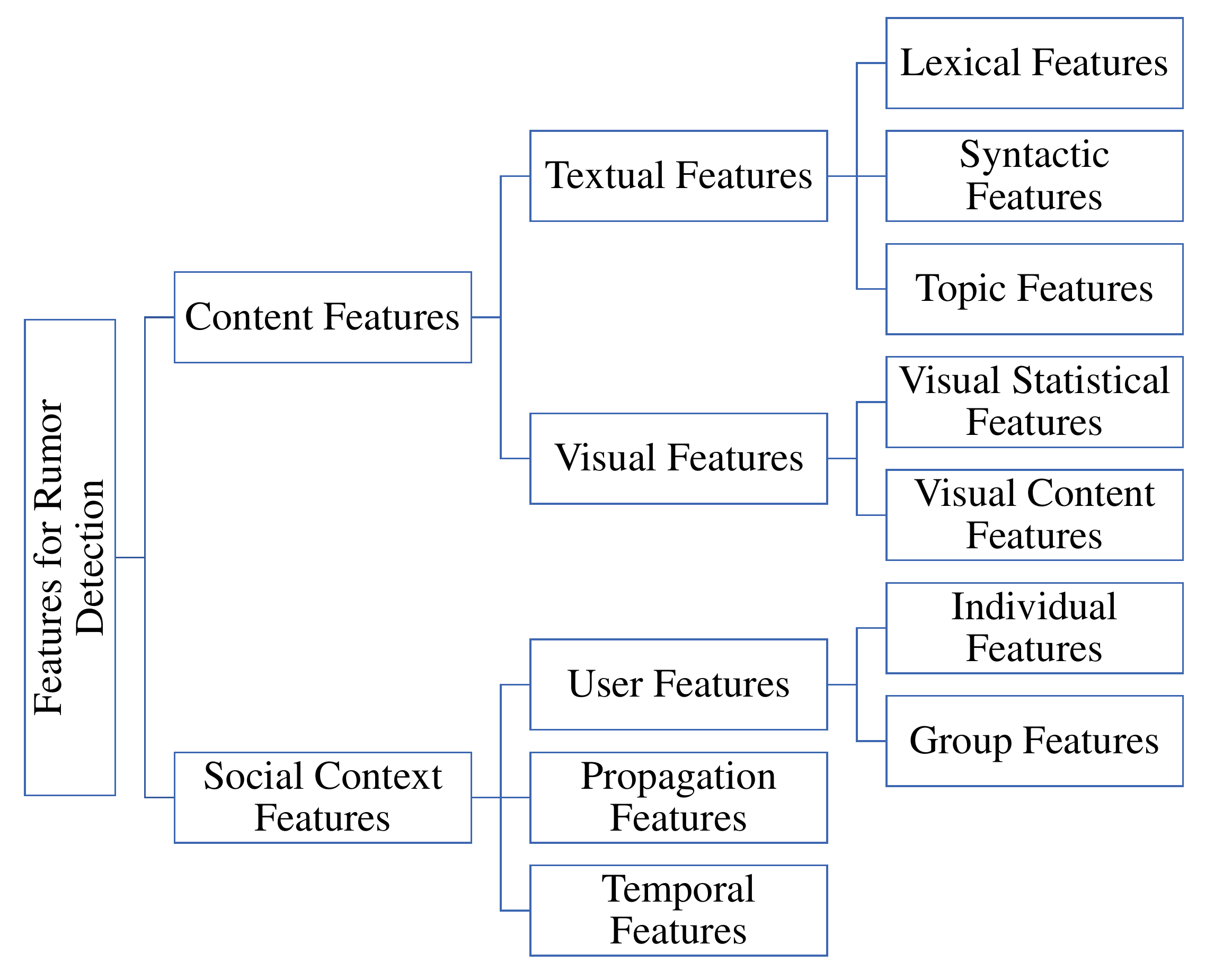}
\caption{Main categories of features used for rumor detection on microblogs.
%\xr{Append references to each feature, say Lexical Feature, Smith et al., 1998, to make the figure more informative.}
}
\label{fig_features}
\end{figure}

In the scope of social media, a news story $e$ contains rich multi-modal resources as its content and associated with certain social context during its spreading, as shown in Figure \ref{fig_example}. Features come from two main aspects of a story: its content and its social context, see Figure \ref{fig_features}. 

The content of a news story contains plenty of information, so we define content features are those features extracted from the text and image. The social context reflects the relationship among different users and describe the propagating process of a rumor, so we define social context features are those features extracted from the user behavior and the propagation network. Two types of features are going to be detailed in Section \ref{ssec:content-feat} and \ref{ssec:context-feat}, respectively.

\subsection{Content Features} \label{ssec:content-feat}

For a news story $e$, the message set $M$ describes all the key information of it, which has content from three aspects:
\begin{itemize}
  \item Textual content: a text describing the news event. They provide details of the event and may contain certain opinions or sentiments towards the story.
  \item Images/Videos: sometimes a message will provide visual material to support its story.
  \item Other content: the specific communication style of social media would provide other informative content, such as hashtag topics(\#), user references(@), links to outer sources, and emoj icons, \emph{etc}.
\end{itemize}

Various content features are extracted from above primitive content, which can be categorized into two main types: textual features and visual features.

\subsubsection{Textual Features}
Compared with non-rumors, rumor texts are fabricated to mislead the public. Aiming to arouse much attention and stimulate the public mood, rumor texts tend to have certain patterns in contrast to non-rumor ones. For example, after analyzing a large amount of tweet streams, Zhao \emph{et al.} \cite{zhao2015enquiring} discover two types of language patterns in rumors: the correction type and the enquiry type. Thus, it is possible to extract features from the textual content, which can characterize rumors.

From different levels granularity of the language structure, various textual features are extracted to represent a rumor from the aspects of words, sentences, messages, topics and events.

General textual features are derived from the linguistics of a text, which are widely used in many natural language understanding tasks. Three categories of general textual features are commonly used: lexical features, syntactic features, and topic features.

\textbf{Lexical features} are features extracted at the word-level of a rumor, which could be statistics of words, lexical rumor patterns or sentimental lexicons.

Castillo \emph{et al.} compute some statistics of rumor texts based on the words it contains, including the total number of words and characters, the number of distinctive words and the average length of word in a rumor, \emph{etc}\cite{castillo2011information}. Apart from simple statistical features, some work extract interesting lexical words by examining their semantics. Kwon \emph{et al.} propose features that describe the fraction of tweets containing the first person pronoun \cite{kwon2013prominent}. Yang \emph{et al.} focus on whether the message includes a URL pointing to an external source. 

According to the study by Zhao \emph{et al.}\cite{zhao2015enquiring}, the informative part of rumor messages can be of the following two types: inquiries of the event situation (verification / confirmation questions) and corrections / disputes of the event. They detect the inquiry and correction patterns of rumor messages through supervised feature selection on a set of labeled messages. Specifically, the uni-grams, bi-grams, and tri-grams lexical parts are extracted from these messages after standard word segmentation. Frequency features (tf) of each extracted word or phrase are then calculated. Following that, Chi-Squared test and information gain ratio method are utilized to select prominent patterns. The feature selection methods rank features based on their ability to distinguish rumor tweets from non-rumor ones. From the ranked list of features, human experts select event-independent phrases as final lexical patterns for rumors. Finally, the lexical word patterns are discovered, including verification patterns ( ``is it real/fake?", ``need more evidence/confirmation", \emph{etc.}) and correction patterns ( ``rumor/false rumor", ``spreading/ fabricating rumor", \emph{etc.}). Once discovered these lexical features can be used to filter out rumors in real time, such as detect rumors in live tweet streams \cite{zhao2015enquiring}\cite{jinImage} or analyzing online rumors at a large scale\cite{jin2017Rumor}.

Lexical words expressing specific semantics or sentiments are also very important clues to characterize the text. In \cite{castillo2011information}, emotional marks( question mark and exclamation mark) and emotion icons are counted as textual features. 
  
  %The occurrences of popular named entities of people, location and organization within a rumor event are unutilized as lexical features in [].%

  In \cite{kwon2013prominent}, many sentimental lexical features are proposed based on sentiment dictionary.   Specifically, they utilize a sentiment tool called the Linguistic Inquiry and Word Count (LIWC) to count words in psychologically meaningful categories. For each input text, this tool provides sentiment analysis results on five major categories and a number of subcategories, such as social, affective, cognitive, perceptual, and biological processes. After comparative study of these features, they find that some categories of sentiments are distinctive features for rumor detection, including positive affect words, cognitive action words and tentative action words.

\textbf{Syntactic features} represent rumors at the sentence level.

  The basic syntactic features are simple statistics of a rumor message, such as the number of keywords, the sentiment score or polarity of the sentence\cite{castillo2011information} and part-of-speech tagging \cite{hassan2010s}.

  To represent a document based on the words it contains, the bag-of-words (BoW) language model is commonly utilized. In this model, each text document is represented as a $v$-dimensional vector, where $v$ is the size of the dictionary of the corpus. Each element in the vector stands for the TF-IDF score of the corresponding word in the text. TF is the term frequency. IDF score is the inverse document frequency, which is calculated on the whole corpus. Some work utilized BoW for rumor detection \cite{ma2017content} or rumor analysis \cite{jin2017Rumor}.

  Recently, semantic embedding algorithms are widely used in many natural language understanding applications. The Word2Vec model \cite{mikolov2013distributed} represents each word in a corpus with a real-valued vector in a common semantic vector space. This algorithm models words based on their semantic analogies. Inspired by its success, some recent rumor detection applications \cite{ma2017content}\cite{jin2017Rumor} also represent texts via word embedding.

\textbf{Topic features} are extracted from the level at the message set, which aim to understand messages and their underlying relations within a corpus.

Wu \emph{et al.} \cite{wu2015false} define a set of topic features to summaries semantics for detecting rumors on Weibo. They train a Latent Dirichlet Allocation (LDA) \cite{blei2003latent} model with an 18-topic distribution on all messages. And each message can belong to one or more topics. They transform the 18-dimensional distribution vector into binary vector by setting the top $k$-highest probability topics to 1 and the rest of topics to 0. The size of $k$ is selected based on the coverage of remained topics.

Jin \emph{et al.} \cite{jin2015} cluster topics based the event a message refers to and extract features both at the message level and the topic level. They assume that messages under a same topic probably have similar credibility values. Under this assumption, they cluster messages into different topics and obtain the topic-level feature by aggregating features on the message level. They claim that this kind of topic-level feature can reduces the impact of noisy data while maintaining most of details on the message level.

Towards a more comprehensive understanding of text on social media, existing work also come up with textual features derived from the traits of social media platform, apart from general textual features. For example, features extracted from outer source links \cite{zhang2015predictors}.

\subsubsection{Visual features}
While most work detects rumors depend on textual content, only some very recent work try to extract features from the visual content of rumors.

Gupta \emph{et al.} \cite{gupta2013faking} make a first attempt to understand the temporal, social reputation and influence patterns for the spreading of fake images on Twitter. They propose a classification model to identify the fake images on Twitter during Hurricane Sandy. Some interesting conclusions are drawn: the original fake images are limited, and 86\% of fake images were re-tweets. These findings could be utilized to design prominent visual features for rumor detection. However, their work is mainly based on traditional textual features.

Aiming to automatically predict whether a tweet that shares multimedia content is fake or real, Boididou \emph{et al.} \cite{verifying2015} proposed the Verifying Multimedia Use task that took place as part of the MediaEval benchmark in 2015 and 2016. This task attracts attentions for verifying images in tweets from the research area.

Visual features are features characterizing the visual content(images or videos) from different aspects. Regarding the methods of feature extraction, visual features can be categorized into three types: visual statistical features, visual content features, and visual embedding features.

\textbf{Visual statistical features} are statistics of images attached in rumors. Similar to the statistical features of textual content, some basic statistics of images proved to be distinctive in separating rumors and non-rumors. 

Gupta \emph{et al.} \cite{gupta2012evaluating} define a feature to record whether the user has a profile image for evaluating the credibility of user. A ``has multimedia" feature is defined in \cite{wu2015false} to record the status of multimedia attachment of a tweet: whether the tweet has any picture, video or audio attached. In \cite{sun2013detecting}, the authors point out that the rumors are more likely to contain outdated images. They propose a time span feature to capture this time delay. In \cite{zhang2015predictors}, the relation of images and health related rumors were studied. The Baidu search engine is deployed to find the original image for calculating this time span between the original image and the current one. According to their results, this feature is quite effective. However, the sparsity of outdated images and the difficulties to search them limit the usage of this feature.

Summarizing existing statistical features and proposing several novel features, seven visual statistical features are presented in \cite{jin2017Novel} from three aspects:

\begin{itemize}
  \item Count: Users can post zero, one or more than one images along with a text content in a tweet. To mark the occurrence of images in rumor messages, they count the total images in a rumor event and the ratio of messages containing at least one or more than one images.
  \item Popularity: Some images are very popular and gain more re-tweets and comments than others in an event. The ratio of the most popular images are calculated to denote this feature.
  \item Image type: Some images have particular type in resolution or style. For example, long images are images with a very large length-to-width ratio. The ratio of these types of images is also counted as a statistical feature.
\end{itemize}

\textbf{Visual content features} are extracted to describe image distributions from a visual perspective, such as visual clarity, diversity and coherence.

In \cite{jin2017Novel}, the authors find that images in rumors and non-rumors are visually distinctive on their distributions. To describe image distributions based on the visual content, five image visual features are proposed in their work: visual clarity score, visual coherence score, visual similarity distribution histogram, visual diversity score, and visual clustering score, which describe image distribution characteristics from different visual aspects.

\begin{itemize}
  \item
  \textbf{Visual Clarity Score} measures the distribution difference between two image sets: one is the image set in a certain news event (event set) and the other is the image set containing images from all events (collection set) \cite{hauff2008improved}. This feature is designed under the assumption that a non-rumor event would contain many original images specifically related to this event, which have a different image distribution in comparison with the overall image set.

  The clarity score is defined as the Kullback-Leibler divergence\cite{kullback1951information} between two language models representing the target image set and all image set, respectively. The bag-of-word image representation is employed to define language models for images. To be specific, local descriptors (e.g. the scale-invariant feature transform (SIFT)\cite{lowe2004distinctive} feature) for each image are extracted and then quantized to form a visual word vocabulary. Each image is represented by words from the vocabulary.

  \item
  \textbf{Visual Coherence Score}  measures how coherent images in a certain news event are \cite{he2008using}\cite{tian2012query}.
 This feature is computed based on visual similarity among images and can reflect relevant relations of images in news events quantitatively.
  More specifically, the average of similarity scores between every two images is computed as the coherence score. In implementation, the similarity between image pairs is calculated based on their GIST feature representations \cite{oliva2001modeling}.

  \item
  \textbf{Visual Similarity Distribution Histogram} describes inter-image similarity in a fine granularity level. It evaluates image distribution with a set of values by quantifying the similarity matrix of each image pair in an event\cite{tian2012query}.
  The visual similarity matrix is obtained by calculating pairwise image similarity in an event. The visual similarity is also computed based on their GIST feature representations. The image similarity matrix is then quantified into an $H$-bin histogram by mapping each element in the matrix into its corresponding bin, which results in a feature vector of $H$ dimensions representing the similarity relations among images.

  \item
  \textbf{Visual Diversity Score} measures the visual difference of the image distribution. Compared with visual coherence score, it computes the image diversity distribution directly and gives more emphasis on representative images.

  The diversity of an image is defined as its minimal difference with the images ranking before it in the whole image set\cite{wang2010}. Here images are ranked according to their popularity on social media, based on the assumption that popular images would have better representation for the news event. The visual diversity score is then calculated as a weighted average of dissimilarity over all images, where top-ranked images have larger weights \cite{deselaers2009jointly}. So that this feature can reduce the impact of unpopular noisy images. Dissimilarity of an image pair is computed as the complementary of its similarity.

  \item
  \textbf{Visual Clustering Score} evaluates the image distribution from a clustering perspective. It is defined as the number of clusters formed by all images in a news event.

  To cluster images, the hierarchical agglomerative clustering (HAC) algorithm \cite{king1967step} is employed, which merges nearest atomic clusters into larger clusters in a bottom-up strategy. HAC does not require the number of clusters to be determined or estimated before clustering. Thus, the number of clusters yielded by HAC reveals the diversity feature of images.

  In implementation, the single-link strategy is used to measure the similarity between two clusters, and the Euclidean distance of image GIST feature vectors is used as the distance measurement. Very small clusters are considered as outliers and removed before calculating the final score \cite{jin2017Novel}.
\end{itemize}

Evaluation of these visual content features on a set of 25,513 images from 146 events shows that images in rumors are more coherentm less diverse and  forming less clusters compared with those in non-rumors. Consequently, these feature can be useful for detecting rumor events.

\subsection{Social context features} \label{ssec:context-feat}
One of the key features of social media is the openness for all kinds of interactions, in comparison with traditional media. Three types of social interactions universally exit on social media:

\begin{itemize}
  \item Interactions among users, such as ``adding friend" and ``following''. This kind of interaction forms the huge and complex underlying network where all information circulates.
  \item Links among multimedia content are formed through tags, hashtag topics or url links. Online contents are organized into sub-groups with this kind of links.
  \item Interactions among users and content, such as ``posting'',``commenting'', ``reposting'', and ``tagging''.
\end{itemize}

Many features are derived from the social connection characteristic of social media on the rumor detection task. Three main types of social features are user feature, propagation features and temporal features.

\textbf{User features} are derived from the user social network. Rumors are created by a few users and spread by many users. Analyzing their characteristics could provide crucial clues for rumor detection. User features could describe the characteristics of a single user or a user group comprised of multiple related users.

\begin{itemize}
  \item
  Individual features are extracted from single user. Common individual features are calculated from a user's  profile, such as ``register time'', ``age'', ``gender''and ``occupation'' \cite{morris2012tweeting}, or user's online actions, such as ``number of followers'', ``number of followees'' and ``number of posted messages'' \cite{castillo2011information}.

  Yang \emph{et al.} \cite{yang2012automatic} proposed two features to mark the user's posting behavior: the ``client'' feature is the software that was used by the user and the ``location'' feature marks whether the message is sent from from where the event happened or not.

  Liu \emph{et al.} \cite{liu2015real} evaluated a user's reliability from the perspective of journalists and proposed user features including the credibility identification, diversity and location of a source user.

  \item Group features are  overall features of a group whose members have certain similar behaviors in the rumor diffusion process \cite{yang2012automatic}. Group features can be generated by aggregating features of single user, such as ``the ratio of verified users'' and ``the average followee count''.
\end{itemize}

\textbf{Propagation features} are derived from the fusion network on which rumors spread. In \cite{castillo2011information}, some statistics of messages' propagation trees, such as the average depth or size of propagation trees, were proposed to capture basic propagation features. The work of Kwon \emph{et al.} \cite{kwon2013prominent} further extended them with 15 structural features extracted from the diffusion network as well as the user friendship network, including the number of nodes and links, the median degree and density of these networks. Wu \emph{et al.} \cite{wu2015false} proposed a concise structure to describe the propagation process for a single message. Yang \emph{et al.} \cite{yang2015exploiting} proposed a set of network features based on network created via the comment providers. In \cite{wang2015detecting}, certain propagation structures are studied to find the distinctive patterns for rumor detection.

\textbf{Temporal features} mark the import time points or the life circle pattern about the diffusion of rumors. In  \cite{kwon2013prominent} message spikes along the time line are modeled to capture the spreading patterns of rumors. Ma \emph{et al.} \cite{ma2015detect} proposed a method for discretizing time stream and capturing the variation of time features. Giasemidis \emph{et al.} \cite{giasemidis2016determining} split every rumor events into 20 time-intervals and extract features for each subset of messages.

Kwon \emph{et al.} studied the stability of features over time \cite{kwon2017rumor}. They found that for rumor detection linguistic and user features are suitable for early stage, while structural and temporal features tend to have good performance in the long-term stage.

\subsection{Classification Methods}
With sufficient features available, many classification methods are proposed in literatures focusing on finding effective features for the rumor detection task.

Most work experimented with more than one classifiers to find the most suitable method, including Decision Tree \cite{castillo2011information}\cite{ma2015detect}\cite{giasemidis2016determining}, Bayesian networks\cite{castillo2011information}, Random Forest\cite{kwon2013prominent}\cite{jin2017Novel}, Logistic Regression \cite{giasemidis2016determining}\cite{zhang2015predictors} and SVM\cite{castillo2011information}\cite{kwon2013prominent}\cite{yang2012automatic}\cite{jin2017Novel}\cite{ma2015detect}.

Some work proposed novel classification algorithms for better aggregating diverse features.  Wu \emph{et al.} \cite{wu2015false}
proposed an SVM with a hybrid kernel technique consisting of random walk kernel \cite{borgwardt2005protein} and an RBF kernel \cite{buhmann2003radial}. The random walk kernel is specifically designed to capture propagation features from the comments tree of a message, while the RBF kernel is applied on content and user features. A two-level classification framework was proposed in \cite{jin2015} to leverage features at the message-level and topic-level. Regarding rumors as anomalies, Chen \emph{et al.}  \cite{chen2016behavior} initially used anomaly detection to classify rumors. 
\section{Propagation-based Approaches} \label{sec:prop}

Hand-crafted features based approaches evaluate each message and event individually. However, some underlying correlations exist among messages and events on social media. One simple observation is that similar messages tend to have the same veracity polarity in an event. The propagation-based approaches are proposed by mining relations among entities and evaluate the credibility of messages and events as a whole. As illustrated in Figure \ref{fig_paradigms}, the paradigm of propagation-based rumor detection commonly has two main steps \cite{gupta2012evaluating}\cite{jin2014news}\cite{jin2016}:

\begin{enumerate}
  \item Constructing a credibility network. Entities involved in rumor detection, such as messages, users, topics or events, are defined as nodes in the network. Each node has an initial credibility value to indicate its confidence of truthfulness. Links among these entities are defined and computed based on their semantic relation or interaction relation on social media.
  \item Credibility propagation. Under some assumptions of node consistency and network smoothness, credibility values are propagated on the constructed network along weighted links until converge, which yields the final credibility evaluation for each entity.
\end{enumerate}

This credibility propagation paradigm is inspired by the work on truth discovery, which aims to find the truth with conflicting information \cite{vydiswaran2011content}. The propagation problem is formed as semi-supervised graph learning task \cite{zhu2002learning}\cite{zhu2003semi}\cite{zhou2004learning}. Compared with direct classification on individual entity, propagation-based approaches can leverage the inter-entity relations and achieve robust results. We give a survey of three typical implementations of propagation-based rumor detection approaches, namely user-message-event network, hierarchical content network and conflicting viewpoints network.

\begin{figure*}[!t]
\centering
\subfigure[User-Message-Event Network]{

\includegraphics[width=1.6in]{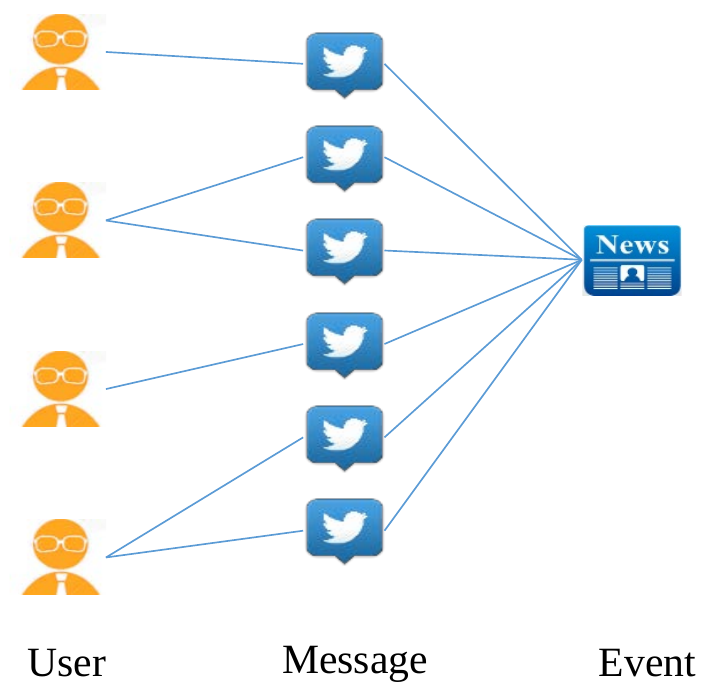}
\label{fig_networka}
}
\subfigure[Hierarchical Content Network]{
\includegraphics[width=1.6in]{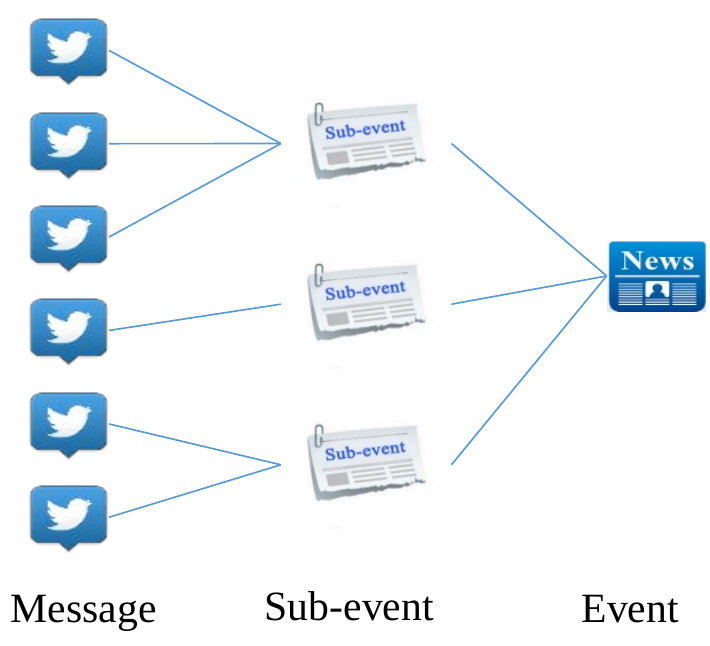}
\label{fig_networkb}
}
\subfigure[Conflicting Viewpoints Network]{

\includegraphics[width=2.7in]{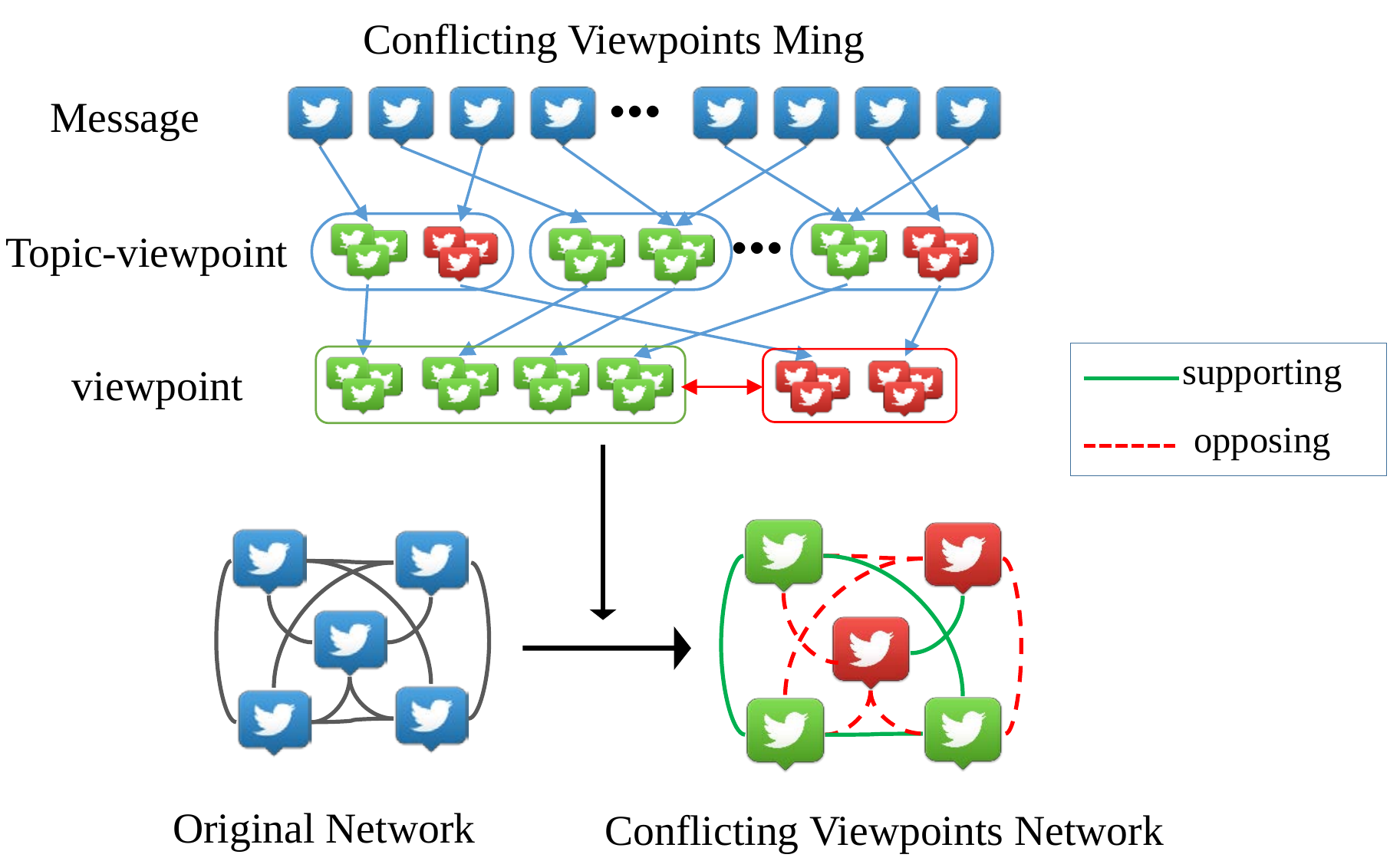}
\label{fig_networkc}
}
\caption{Three different networks for propagation-based rumor detection.}
\label{fig_network}
\end{figure*}

\subsection{User-Message-Event Network}
Gupta \emph{et al.} \cite{gupta2012evaluating} first introduced the propagation method . They constructed a network consists of users, messages and events under two intuitions:
\begin{itemize}
  \item Credible users do not offer credibility to rumor events in general.
  \item Links between credible messages have larger weights than those for rumor messages, as messages in an rumor event don't make coherent claims.
\end{itemize}

\subsubsection{Network structure}
As illustrated in Figure \ref{fig_networka}, the structure of this network is formed as followings:
\begin{itemize}
  \item Each user is linked to a message if puts up that message.
  \item Each message is linked to an event if it belongs to that event (i.e. containing the same keywords as the event).
  \item Messages are linked with each other to denote the inter-message relationship.
  \item Events are linked with other events to denote inter-event relationship.
\end{itemize}

The inter-message links denote how a message would influence other messages. These influence is computed as semantic similarity between two messages in \cite{gupta2012evaluating}. The ratio of shared unigrams is calculated as the link's weight to represent the influence degree between two linked messages.  Similarly, the links between events are computed as the shared unigrams in the event keywords representing two events. As for the weights of the rest two types of links: links from users to messages and links from messages to events, are not defined and set as 1 by default.

\subsubsection{Credibility propagation}

The initial credibility values of each messages are obtained from the results of a feature-based classifier, similar to that of \cite{castillo2011information}. Then they are propagated on this network using PageRank-like iterations. At each iteration, the credibility valuses are updated as follows:
\begin{itemize}
  \item For each message, its credibility value is affected by three aspects: the user and event it links to and other linked messages. Contributions from these aspects are weighted to make sure that they are comparable when updating its credibility.
  \item For each event, its credibility value is affected by two aspects: the messages it links to and other linked events. Therefore, its credibility value is updated accordingly.
  \item For each user, the average credibilities of all the messages it links to is computed for updating.
\end{itemize}

After several iterations, this algorithm would converge and produce the final predictions for each entity.

%The authors report that the credibility propagation can improve rumor detection on two tweet event set from $\sim$74\% to $\sim$76\%. And with optimization on the event-level, the accuracy can be further boosted to over 85\%. %

\subsection{Hierarchical Content Network}
Inspired by the idea of linking entities altogether and leveraging inter-entity implications for credibility propagation, Jin \emph{et al.} \cite{jin2014news} proposed a three-layer credibility network constructed from different semantic levels of content of an event. They firstly pointed out that the involving of users in the network in \cite{gupta2012evaluating} is less convincible. Based on their observation, many users spread rumors unintentionally on social media and even credible users would be misguided to spread rumors. They also found that a news event, as a whole, contains both real and fake information in many occasions. Thus, without deeper analysis of its components, it is hard to get a convincing evaluation for the event. Motivated by these observations, they aimed to minimize users' influence and focus on an event's deeper semantic relations by proposing a three-layer credibility propagation network.

\subsubsection{Network structure}

The hierarchical content network has three layers of entities: message layer, sub-event layer and event layer (Figure \ref{fig_networkb}). They are all content-based, and have direct relations with news credibility. The sub-event layer is initially introduced to capture the deeper sematic information within an event. Sub-events are various point of views of an event which are message clusters representing major parts or topics of an event. To be specific, the network is formed as followings: a message is linked to a sub-vent if it is clustered into that sub-events; a sub-event is linked to the event; all the messages are linked with each other so are the sub-events.

Similar to \cite{gupta2012evaluating}, the four types of inter-entity links denote how entities would influence each other on credibility evaluation. In \cite{jin2014news}, the link weights are mostly determined by two facts: the semantic similarity and the social importance. By assuming semantically similar entities would have similar credibility values, the sematic similarity scores are computed between entity pairs based on their word representation. Meanwhile, the social importance score, which is calculated from the number of re-tweets or comments gained by an entity, is incorporated to give emphasis on more popular content in a sub-event or event.

\subsubsection{Credibility propagation}
Under the assumption that entities with large link weight between them should have similar credibility values, the credibility propagation problem is formulated as a graph optimization problem. The authors gave a loss function which are constraints ensure the propagation should not change too much between entities with large link weight or from the entities' initial values. They then deduced the iterative solution for optimizing the loss function with the gradient descent method. The authors claimed that their formulation of the credibility propagation is a convex problem which ensures that they can provide the globally optimal solution with an iterative algorithm. Compared with the institutive propagation method in \cite{gupta2012evaluating}, this is a theoretical improvement.

\subsection{Conflicting Viewpoints Network}

According to \cite{jin2016}, there exist two kinds of relations between messages on microblogs. One relation is supporting, where messages expressing the same viewpoint support each others credibility. The other relation is opposing, where messages expressing conflicting viewpoints decrease each other's credibility. As microblogs are open media platforms, people can post their skeptical and even opposing responses after they read a news event. These opposing voices would arise against the news along with original supportive voices in the case of rumors, which are very crucial components for evaluating the truthfulness of news events.

\subsubsection{Network structure}
Based on this observation, the authors proposed a credibility network with both supporting and opposing relations by exploiting the conflicting viewpoints in microblogs (Figure \ref{fig_networkc}).

\begin{itemize}
  \item Conflicting viewpoints mining. Conflicting viewpoints are mined through a joint topic-viewpoint model \cite{trabelsi2014mining}. This model represents each message as a mixture of topics and a mixture of viewpoints for each topic. These topic-viewpoint pairs are then clustered under constraints to form the final conflicting viewpoints.
  \item Link definition. The link between any two messages are computed based on the results of mined conflicting viewpoints. The wight of the link is computed as the distance between the probability representations of the two messages from the topic model. And the polarity of the link is defined from the viewpoints clustering: messages have the same viewpoints form the positive link, otherwise they form the negative link.
  \item Network construction. All messages in an event are linked with each other. Links between them are computed as above.
\end{itemize}

\subsubsection{Credibility propagation}
Similar to that in \cite{jin2014news}, the credibility propagation on this network is also defined as a graph optimization problem. To deal with the negative links in the network, the authors propose a loss function which can ensure that messages with supporting relations have similar credibility values and messages with opposing relations have opposite credibility values or values both close to zero. Given the loss function, they also derived the optimal solution to it.

\section{Deep Neural Networks Approaches}
Compared to the traditional classifiers, deep neural networks (DNN) have demonstrated clear advantages for many machine learning problems, such as object detection, sentiment classification and voice recognition. DNN based methods aim to learn deep representation of rumor data automatically. Based on the different structures of neural networks, the neural network methods can be further classified into the two categories:

\begin{itemize}
  \item
  Recurrent Neural Networks: Based on the structure of RNN, this kind of approach models the rumor data as sequential data. The key point is that the connections between units in an RNN form a direct cycle and create an internal state of the network that might allow it to capture the dynamic temporal signals characteristic of rumor diffusion \cite{ma2016detecting,chen2017call,ruchansky2017csi,jin2017multimodal}.
  \item
   Convolutional Neural Networks:  CNN is composed by stacked convolutional and pooling layers, the structure of which help model significant semantic features. The CNN based methods\cite{yuconvolutional} \cite{nguyen2017early} assume that CNN can not only automatically extract local-global significant features from an input instance but reveal those high-level interactions. 

\end{itemize}

We now introduce some representative works for each category.

\subsection{RNN-based methods}

Ma \emph{et al.} first apply recurrent neural network to detect rumors. They observe that a rumor event consists of an original post and a group of related posts including reposts and comments, which create a continues stream of posts. Thus, they model the rumor data as a variable length time series. However, a rumor event consists of tens of thousands of posts, so they batch posts into time intervals and treat them as a single unit in a time series that is then modeled using an RNN sequence. In each interval, they use the $tf*idf$ values of the top-K terms in the vocabulary as input. Their model aims to learn both  temporal and textual representations from the rumor data under supervision and extensive experiments demonstrate that their model achieves outstanding performance compared to those work utilizing hand-crafted features.

Some malicious words in the content may be strongly related to the rumor category. To better understand what words the model pay more attention to, Chen \emph{et al.} leverage attention mechanism in their model\cite{chen2017call}. One assumption of their work is that textual features of the rumor data may change their importance with time and it is crucial to determine which of them are more important to the detection task. Similar to \cite{ma2016detecting}, they first batch posts into time intervals and use $tf*idf$ as the representation of the input. In each timestep, the hidden state will be allocated a weight parameter to measure its importance and contribution to the results. The performance of experiments demonstrate the efficiency of attention mechanism and show that most words connected with the event itself are given less attention weight than words expressing users' doubting, esquiring and anger caused by the rumor, while ignore unrelated words.

Ruchansky \emph{et al.} focus on three characteristics of the rumor data\cite{ruchansky2017csi}: the text of an article, the user response it receives, and the source users promoting it. These characteristics represent different aspects of the rumor data, and it is also challenging to detect rumors based on one of them. Thus, they propose a hybrid model(CSI) that combines all three characteristics for a more accurate and automated prediction. The model is composed of three modules: Capture, Score, and Integrate. The first module is based on the response and text; it uses a Recurrent Neural Network to capture the temporal pattern of user activity on a given article. The second module learns the source characteristic based on the behavior of users, and a user is represented by a vector. In the third module, the result of the first two modules are integrated to a vector which is used to classify an article as fake or not. Aside from accurate prediction, the CSI model also produces latent representations of both users and articles that can be used for separate analysis.

Jin \emph{et al.} not only leverage the textual information but also visual and social information and propose a multimodal fusion based model. An increasing number of users are using images and videos to post news in addition to texts. Therefore, for given a post, its text and social context are first fused with an LSTM unit. The joint representation are then fused with visual features extracted from pre-trained deep VGG-19. The output of the LSTM at each time step is employed as the neuron-level attention to coordinate visual features during the fusion. Extensive experiments conducted on Weibo and Twitter datasets demonstrate that their model can effectively detect rumors based on multimedia content, in comparison with existing feature-based methods and several multimodal fusion methods based on neural networks.

\subsection{CNN-based methods}

Yu \emph{et al.} find that the RNN is not qualified to the early detection tasks with limited inputting data and has a bias towards the latest elements of input sequence \cite{yuconvolutional}. To address those issues, they propose a convolutional neural network based approach for rumor detection. Specifically, they put forward a method to split every rumor event into several phases. Subsequently, all events are split into several groups of microblog posts. And representation of each group is learnt though doc2vec, so an input sequence of CAMI is formed of a group of vectors. Finally, the vectors are fed to a two-layer convolutional neural networks, obtaining the final results of two-class classification. Their model can extract significant features from an input instance and achieve high performance on the two open dataset. 

Nguyen \emph{et al.} focus on detect rumors at the early stage of rumor propagation, who proposed a CNN + RNN based model. In the unified model, the part of CNN is applied for representation. Each tweet is first formed of a set of word embeddings which are learnt jointly in the model training process. Specifically, the model utilizes CNN to extract a sequence of higher-level phrase representations to learn the hidden representations of individual rumor-related tweets. Then the part of RNN is used to process the time series obtained by CNN. By using limited rumor data, extensive experiments show the good performance of their model within the very first hours during the propagation of a rumor.

\section{Datasets}
Since different papers report experimental results utilizing different datasets, to compare methods fairly on a reasonable scale, a few benchmark datasets have recently been proposed. We review the following representative datasets because they were collected from real-world social media platforms and have been widely used in rumor detection. 

KWON dataset. The KWON dataset \cite{kwon2013prominent}, released in 2013, consists of 47 events of rumor and 55 events of non-rumor collected from Twitter, each events contains 60 tweets at least. To ensure all the events are valid, each event is labeled by four participants and the dataset only contains those events that are evaluated by at least four participants and had the majority agreement.

MediaEval dataset. The MediaEval dataset is released by The Verifying Multimedia Use task at MediaEval\cite{boididou2014challenges}, which aims to detect false multimedia content on Twitter. It contains 9000 tweets of rumor and 6000 non-rumor tweets related to 17 events in the development set, and 2000 tweets related to 35 events in the test set.

RUMDECT dataset. The RUMDECT dataset \cite{ma2016detecting}, released in 2016, consists of two types of data from Weibo and Twitter, respectively. For Weibo data, a group of known rumors are first collected from Sina community management center and another group of non-rumor events are crawled from the posts of general threads. Therefore, the Weibo data contains 2313 rumors and 2351 non-rumors. For the Twitter data, 778 reported events during March-December 2015 are collected from Snopes.com. For each event, the keywords from its Snopes URL are applied as query to search related posts from Twitter. In order to balance the two classes, some non-rumor events are added from two another datasets \cite{castillo2011information,kwon2013prominent}. The resulting twitter data contains 498 rumor events and 494 non-rumor events.

RUMOUREVAL dataset. The RUMOUREVAL dataset is produced for RumourEval 2017 \cite{derczynski2017semeval}, a shared subtask in SemEval 2017. The training set consists of 297 rumourous threads collected for 8 events, which include 297 source and 4,222 reply tweets. The testing set includes, in total, 1,080 tweets, 28 of which are source tweets and 1,052 replies.

MULTI dataset. The MULTI dataset \cite{jin2017multimodal} released in 2017, includes 4749 posts of rumor and 4779 posts of non-rumor collected from the official rumor debunking system of Weibo. Unlike the previous datasets, the MULTI dataset is the first dataset that focuses on leveraging multimodal contents to detect rumors on Weibo platform, which contains not only textual information but also visual information.

\begin{table}\centering
\label{tab:datasets}
\caption{Statistics on the datasets}
\ra{1.3}
\begin{tabular}
{@{}llll@{}}
\toprule
Dataset & Data source & Rumor & Non-rumor\\
\midrule
KWON & Twitter & 47 & 55  \\
MediaEval & Twitter & 9000 & 6000  \\
RUMDECT & Twitter \& Weibo &498 & 494 \\
RUMOURREVAL & Twitter & 145 & 74 \\
MULTI & Weibo & 4749 & 4779 \\
\bottomrule
\end{tabular}
\end{table}

Table \uppercase\expandafter{\romannumeral1} summarizes the datasets used for rumor detection. Based on their statistic information, we can observe that each dataset either consists of a set of events or a set of messages. Datasets containing rumor events are suited for classifying a wide spread event on social media platforms, while datasets composed by rumor messages are suited for classifying a single message.

\section{Conclusion and Perspectives}

Rumors spreading on social media could severely influence people' daily life. However, most applications still rely on manual efforts, either by human experts or by common user, to combat the ever-increasing rumors. Under this circumstances, the research on automatic rumor detection attracts more and more attention. In this survey, we attempt to provide a comprehensive survey for exiting studies towards the development of automatic rumor detection approaches. Moreover, we have categorized the existing work into three paradigms, hand-crafted features based approach, propagation-based approach and neural networks approach. We start from introducing the features of rumors and then we analyze the principles of the existing classification algorithms. 

Three different paradigms for rumor detection are elaborately described: the feature-based classification approach, the credibility propagation approach and the neural networks approaches. For each method in these paradigm, it aims to tow goals: one is to extract prominent features to representing the multimedia content comprise rumors and the social context generated by rumors on social network, the other goal is to build robust machine learning algorithms to separate rumors from common stories. 
%\xr{this paragraph requires rewrite. Any INSPIRING conclusions about the three paradigms? Current conclusion will not satisfy the TKDE reviewers}

Moreover, we introduce several public datasets and give the statistic details for each of them. All the datasets are collected from Weibo or Twitter platform. Each sample in event based dataset refers to a group of messages with one topic, while the sample in message based dataset refers to a single message . We can observe an encouraging trend that some works, begin to build large datasets. But in most works, the datasets
are still not large enough. A main difficulty in collecting dataset is that rumor are those news debunked by authorities, the number of which is less than that of nonrumor samples. 
%\xr{this paragraph requires rewrite. Not good enough. Such a conclusion will not satisfy the TKDE reviewers}

Apart from finding more efficient features or algorithms, we summary the four main challenges for future studies:

Early detection. The life circle of a story propagating on social networks is quite short, some studies suggested it is less than three days. What's more, rumors become viral in seconds or minutes \cite{friggeri2014rumor}. It is crucial to detect rumors at their very early stage. However, most existing studies detect rumors by assuming they have all content in all life time of the rumor. The resource on the beginning of a rumor is such limited that it is very challenging to detect it at the early stage. Although Yu \emph{et al.} propose a method to solve this problem, the performance is unable to meet the need of early detection. 
 
 Explanatory detection. Exiting rumor detection methods only give a final decision of whether a story is a rumor. Little information is revealed why they make the decision. However, unlike simple object classification, finding the evidences supporting the decision would be beneficial in debunking the rumor and prevent its further spreading. The explanatory rumor detection requires algorithms to monitor much more closely into every components in rumors, which is a challenge to be resolved.
  
  Long text rumor detection. The objects of rumor detection methods at present are short text spread on social media. However, more and more long text news such as blogs and passages are generated on some online communities, which arise a growing demand for verification. Different from short text rumors, long text rumors have affluent semantic information which poses an obvious obstacle for comprehensive understanding. Moreover, in most cases, only part of a given long text rumor contains false information, while the rest of which is true. So it is unfair to classify the whole passage as a rumor or not. More efforts should be addressed not only on improving the performance for detecting long text rumor but also on pointing out the precise position of false information in a passage.
 
Multimodal rumor detection. The fact that an increasing number of rumors consisting of multimodal data are propagating on social media brings difficulties for traditional detection methods. Therefore, analyzing the relationships among data with multiple modalities and developing advanced fusion based models to utilize these data can be the key to detect rumors in more complex scenarios. Though Jin \emph{et al.} proposed a fusion based model to detection rumors with multiple modalities\cite{jin2017multimodal}, the complex relationship among different modalities should be modeled more accurately.

\ifCLASSOPTIONcaptionsoff
  
\fi

\bibliographystyle{IEEEtran}

\bibliography{survey}

% Generated by IEEEtran.bst, version: 1.14 (2015/08/26)
\begin{thebibliography}{10}
\providecommand{\url}[1]{#1}
\csname url@samestyle\endcsname
\providecommand{\newblock}{\relax}
\providecommand{\bibinfo}[2]{#2}
\providecommand{\BIBentrySTDinterwordspacing}{\spaceskip=0pt\relax}
\providecommand{\BIBentryALTinterwordstretchfactor}{4}
\providecommand{\BIBentryALTinterwordspacing}{\spaceskip=\fontdimen2\font plus
\BIBentryALTinterwordstretchfactor\fontdimen3\font minus
  \fontdimen4\font\relax}
\providecommand{\BIBforeignlanguage}[2]{{%
\expandafter\ifx\csname l@#1\endcsname\relax
\typeout{** WARNING: IEEEtran.bst: No hyphenation pattern has been}%
\typeout{** loaded for the language `#1'. Using the pattern for}%
\typeout{** the default language instead.}%
\else
\language=\csname l@#1\endcsname
\fi
#2}}
\providecommand{\BIBdecl}{\relax}
\BIBdecl

\bibitem{Gottfried2016}
J.~Gottfried and E.~Shearer, ``News use across social media platforms 2016,''
  \emph{Pew Research Center}, 2016.

\bibitem{zhao2015enquiring}
Z.~Zhao, P.~Resnick, and Q.~Mei, ``Enquiring minds: Early detection of rumors
  in social media from enquiry posts,'' in \emph{Proceedings of the 24th
  International Conference on World Wide Web}.\hskip 1em plus 0.5em minus
  0.4em\relax International World Wide Web Conferences Steering Committee,
  2015, pp. 1395--1405.

\bibitem{friggeri2014rumor}
A.~Friggeri, L.~A. Adamic, D.~Eckles, and J.~Cheng, ``Rumor cascades,'' in
  \emph{Proceedings of the Eighth International AAAI Conference on Weblogs and
  Social Media}, 2014.

\bibitem{jin2017Rumor}
Z.~Jin, J.~Cao, H.~Guo, Y.~Zhang, Y.~Wang, and J.~Luo, ``Detection and analysis
  of 2016 {US} presidential election related rumors on twitter,'' in
  \emph{Social, Cultural, and Behavioral Modeling - 10th International
  Conference, SBP-BRiMS 2017, Washington, DC, USA, July 5-8, 2017,
  Proceedings}, 2017, pp. 14--24.

\bibitem{Domm2013}
P.~Domm, ``False rumor of explosion at white house causes stocks to briefly
  plunge; ap confirms its twitter feed was hacked,'' 2013.

\bibitem{jin2014news}
Z.~Jin, J.~Cao, Y.-G. Jiang, and Y.~Zhang, ``News credibility evaluation on
  microblog with a hierarchical propagation model,'' in \emph{2014 IEEE
  International Conference on Data Mining (ICDM)}.\hskip 1em plus 0.5em minus
  0.4em\relax IEEE, 2014, pp. 230--239.

\bibitem{Facebook2016}
``Facebook to begin flagging fake news in response to mounting criticism,''
  \url{https://www.theguardian.com/technology/2016/dec/15/facebook-flag-fake-news-fact-check},
  2016, [Online; Accessed 15-December-2016].

\bibitem{Gupta2014TweetCred}
A.~Gupta, P.~Kumaraguru, C.~Castillo, and P.~Meier, \emph{TweetCred: Real-Time
  Credibility Assessment of Content on Twitter}.\hskip 1em plus 0.5em minus
  0.4em\relax Springer International Publishing, 2014, pp. 228--243.

\bibitem{ma2016detecting}
J.~Ma, W.~Gao, P.~Mitra, S.~Kwon, B.~J. Jansen, K.-F. Wong, and M.~Cha,
  ``Detecting rumors from microblogs with recurrent neural networks,'' in
  \emph{Proceedings of IJCAI}, 2016.

\bibitem{Wang2012}
M.~Wang, B.~Ni, X.-S. Hua, and T.-S. Chua, ``Assistive tagging: A survey of
  multimedia tagging with human-computer joint exploration,'' \emph{ACM Comput.
  Surv.}, vol.~44, no.~4, pp. 25:1--25:24, Sep. 2012.

\bibitem{castillo2011information}
C.~Castillo, M.~Mendoza, and B.~Poblete, ``Information credibility on
  twitter,'' in \emph{Proceedings of the 20th international conference on World
  Wide Web (WWW)}.\hskip 1em plus 0.5em minus 0.4em\relax ACM, 2011, pp.
  675--684.

\bibitem{yang2012automatic}
F.~Yang, Y.~Liu, X.~Yu, and M.~Yang, ``Automatic detection of rumor on sina
  weibo,'' in \emph{Proceedings of the ACM SIGKDD Workshop on Mining Data
  Semantics}.\hskip 1em plus 0.5em minus 0.4em\relax ACM, 2012, pp. 1--7.

\bibitem{kwon2013prominent}
S.~Kwon, M.~Cha, K.~Jung, W.~Chen, and Y.~Wang, ``Prominent features of rumor
  propagation in online social media,'' in \emph{Data Mining (ICDM), 2013 IEEE
  13th International Conference on}.\hskip 1em plus 0.5em minus 0.4em\relax
  IEEE, 2013, pp. 1103--1108.

\bibitem{sun2013detecting}
S.~Sun, H.~Liu, J.~He, and X.~Du, ``Detecting event rumors on sina weibo
  automatically,'' in \emph{Web Technologies and Applications}.\hskip 1em plus
  0.5em minus 0.4em\relax Springer, 2013, pp. 120--131.

\bibitem{wu2015false}
K.~Wu, S.~Yang, and K.~Q. Zhu, ``False rumors detection on sina weibo by
  propagation structures,'' in \emph{IEEE International Conference on Data
  Engineering, ICDE}, 2015.

\bibitem{jin2015}
Z.~Jin, J.~Cao, Y.~Zhang, and Z.~Yongdong, ``Mcg-ict at mediaeval 2015:
  Verifying multimedia use with a two-level classification model,'' in
  \emph{Proceedings of the MediaEval 2015 Multimedia Benchmark Workshop}, 2015.

\bibitem{gupta2012evaluating}
M.~Gupta, P.~Zhao, and J.~Han, ``Evaluating event credibility on twitter,'' in
  \emph{Proceedings of the SIAM International Conference on Data Mining}.\hskip
  1em plus 0.5em minus 0.4em\relax Society for Industrial and Applied
  Mathematics, 2012, p. 153.

\bibitem{jin2016}
Z.~Jin, J.~Cao, Y.~Zhang, and J.~Luo, ``News verification by exploiting
  conflicting social viewpoints in microblogs,'' in \emph{Proceedings of the
  Thirtieth {AAAI} Conference on Artificial Intelligence, February 12-17, 2016,
  Phoenix, Arizona, {USA.}}, 2016.

\bibitem{chen2017call}
T.~Chen, L.~Wu, X.~Li, J.~Zhang, H.~Yin, and Y.~Wang, ``Call attention to
  rumors: Deep attention based recurrent neural networks for early rumor
  detection,'' \emph{arXiv preprint arXiv:1704.05973}, 2017.

\bibitem{ruchansky2017csi}
N.~Ruchansky, S.~Seo, and Y.~Liu, ``Csi: A hybrid deep model for fake news
  detection.''\hskip 1em plus 0.5em minus 0.4em\relax CIKM, 2017.

\bibitem{jin2017multimodal}
Z.~Jin, J.~Cao, H.~Guo, Y.~Zhang, and J.~Luo, ``Multimodal fusion with
  recurrent neural networks for rumor detection on microblogs,'' in
  \emph{Proceedings of the 2017 {ACM} on Multimedia Conference, {MM} 2017,
  Mountain View, CA, USA, October 23-27, 2017}, 2017, pp. 795--816.

\bibitem{yuconvolutional}
F.~Yu, Q.~Liu, S.~Wu, L.~Wang, and T.~Tan, ``A convolutional approach for
  misinformation identification.''

\bibitem{nguyen2017early}
T.~N. Nguyen, C.~Li, and C.~Nieder{\'e}e, ``On early-stage debunking rumors on
  twitter: Leveraging the wisdom of weak learners,'' in \emph{International
  Conference on Social Informatics}.\hskip 1em plus 0.5em minus 0.4em\relax
  Springer, 2017, pp. 141--158.

\bibitem{zubiaga2017detection}
A.~Zubiaga, A.~Aker, K.~Bontcheva, M.~Liakata, and R.~Procter, ``Detection and
  resolution of rumours in social media: A survey,'' \emph{arXiv preprint
  arXiv:1704.00656}, 2017.

\bibitem{boididou2017verifying}
C.~Boididou, S.~E. Middleton, Z.~Jin, S.~Papadopoulos, D.-T. Dang-Nguyen,
  G.~Boato, and Y.~Kompatsiaris, ``Verifying information with multimedia
  content on twitter,'' \emph{Multimedia Tools and Applications}, pp. 1--27,
  2017.

\bibitem{difonzo2007rumor}
N.~DiFonzo and P.~Bordia, \emph{Rumor psychology: Social and organizational
  approaches}.\hskip 1em plus 0.5em minus 0.4em\relax American Psychological
  Association Washington, DC, 2007, vol.~1.

\bibitem{allport1947psychology}
G.~W. Allport and L.~Postman, ``The psychology of rumor.'' 1947.

\bibitem{alexander2014social}
D.~E. Alexander, ``Social media in disaster risk reduction and crisis
  management,'' \emph{Science and engineering ethics}, vol.~20, no.~3, pp.
  717--733, 2014.

\bibitem{morris2012tweeting}
M.~R. Morris, S.~Counts, A.~Roseway, A.~Hoff, and J.~Schwarz, ``Tweeting is
  believing?: understanding microblog credibility perceptions,'' in
  \emph{Proceedings of the ACM 2012 conference on Computer Supported
  Cooperative Work}.\hskip 1em plus 0.5em minus 0.4em\relax ACM, 2012, pp.
  441--450.

\bibitem{jin2017Novel}
Z.~Jin, J.~Cao, Y.~Zhang, J.~Zhou, and Q.~Tian, ``Novel visual and statistical
  image features for microblogs news verification,'' \emph{{IEEE} Trans.
  Multimedia}, vol.~19, no.~3, pp. 598--608, 2017.

\bibitem{jinImage}
Z.~Jin, J.~Cao, J.~Luo, and Y.~Zhang, ``Image credibility analysis with
  effective domain transferred deep networks,'' \emph{CoRR}, vol.
  abs/1611.05328, 2016.

\bibitem{hassan2010s}
A.~Hassan, V.~Qazvinian, and D.~Radev, ``What's with the attitude?: identifying
  sentences with attitude in online discussions,'' in \emph{Proceedings of the
  2010 Conference on Empirical Methods in Natural Language Processing}.\hskip
  1em plus 0.5em minus 0.4em\relax Association for Computational Linguistics,
  2010, pp. 1245--1255.

\bibitem{ma2017content}
B.~Ma, D.~Lin, and D.~Cao, ``Content representation for microblog rumor
  detection,'' in \emph{Advances in Computational Intelligence Systems}.\hskip
  1em plus 0.5em minus 0.4em\relax Springer, 2017, pp. 245--251.

\bibitem{mikolov2013distributed}
T.~Mikolov, I.~Sutskever, K.~Chen, G.~S. Corrado, and J.~Dean, ``Distributed
  representations of words and phrases and their compositionality,'' in
  \emph{Advances in neural information processing systems}, 2013, pp.
  3111--3119.

\bibitem{blei2003latent}
D.~M. Blei, A.~Y. Ng, and M.~I. Jordan, ``Latent dirichlet allocation,''
  \emph{Journal of machine Learning research}, vol.~3, no. Jan, pp. 993--1022,
  2003.

\bibitem{zhang2015predictors}
Z.~Zhang, Z.~Zhang, and H.~Li, ``Predictors of the authenticity of internet
  health rumours,'' \emph{Health Information \& Libraries Journal}, vol.~32,
  no.~3, pp. 195--205, 2015.

\bibitem{gupta2013faking}
A.~Gupta, H.~Lamba, P.~Kumaraguru, and A.~Joshi, ``Faking sandy: characterizing
  and identifying fake images on twitter during hurricane sandy,'' in
  \emph{Proceedings of the 22nd international conference on World Wide Web
  companion}.\hskip 1em plus 0.5em minus 0.4em\relax International World Wide
  Web Conferences Steering Committee, 2013, pp. 729--736.

\bibitem{verifying2015}
C.~Boididou, K.~Andreadou, S.~Papadopoulos, D.-T. Dang-Nguyen, G.~Boato,
  M.~Riegler, and Y.~Kompatsiaris, ``Verifying multimedia use at mediaeval
  2015,'' in \emph{MediaEval 2015 Workshop, Sept. 14-15, 2015, Wurzen,
  Germany}, 2015.

\bibitem{hauff2008improved}
C.~Hauff, V.~Murdock, and R.~Baeza-Yates, ``Improved query difficulty
  prediction for the web,'' in \emph{Proceedings of the 17th ACM conference on
  Information and knowledge management}.\hskip 1em plus 0.5em minus 0.4em\relax
  ACM, 2008, pp. 439--448.

\bibitem{kullback1951information}
S.~Kullback and R.~A. Leibler, ``On information and sufficiency,'' \emph{The
  annals of mathematical statistics}, pp. 79--86, 1951.

\bibitem{lowe2004distinctive}
D.~G. Lowe, ``Distinctive image features from scale-invariant keypoints,''
  \emph{International journal of computer vision}, vol.~60, no.~2, pp. 91--110,
  2004.

\bibitem{he2008using}
J.~He, M.~Larson, and M.~De~Rijke, ``Using coherence-based measures to predict
  query difficulty,'' in \emph{Advances in Information Retrieval}.\hskip 1em
  plus 0.5em minus 0.4em\relax Springer, 2008, pp. 689--694.

\bibitem{tian2012query}
X.~Tian, Y.~Lu, and L.~Yang, ``Query difficulty prediction for web image
  search,'' \emph{IEEE Transactions on Multimedia}, vol.~14, no.~4, pp.
  951--962, 2012.

\bibitem{oliva2001modeling}
A.~Oliva and A.~Torralba, ``Modeling the shape of the scene: A holistic
  representation of the spatial envelope,'' \emph{International journal of
  computer vision}, vol.~42, no.~3, pp. 145--175, 2001.

\bibitem{wang2010}
M.~Wang, K.~Yang, X.~S. Hua, and H.~J. Zhang, ``Towards a relevant and diverse
  search of social images,'' \emph{IEEE Transactions on Multimedia}, vol.~12,
  no.~8, pp. 829--842, Dec 2010.

\bibitem{deselaers2009jointly}
T.~Deselaers, T.~Gass, P.~Dreuw, and H.~Ney, ``Jointly optimising relevance and
  diversity in image retrieval,'' in \emph{Proceedings of the ACM international
  conference on image and video retrieval}.\hskip 1em plus 0.5em minus
  0.4em\relax ACM, 2009, p.~39.

\bibitem{king1967step}
B.~King, ``Step-wise clustering procedures,'' \emph{Journal of the American
  Statistical Association}, vol.~62, no. 317, pp. 86--101, 1967.

\bibitem{liu2015real}
X.~Liu, A.~Nourbakhsh, Q.~Li, R.~Fang, and S.~Shah, ``Real-time rumor debunking
  on twitter,'' in \emph{Proceedings of the 24th ACM International on
  Conference on Information and Knowledge Management}.\hskip 1em plus 0.5em
  minus 0.4em\relax ACM, 2015, pp. 1867--1870.

\bibitem{yang2015exploiting}
Y.~Yang, K.~Niu, and Z.~He, ``Exploiting the topology property of social
  network for rumor detection,'' in \emph{Computer Science and Software
  Engineering (JCSSE), 2015 12th International Joint Conference on}.\hskip 1em
  plus 0.5em minus 0.4em\relax IEEE, 2015, pp. 41--46.

\bibitem{wang2015detecting}
S.~Wang and T.~Terano, ``Detecting rumor patterns in streaming social media,''
  in \emph{Big Data (Big Data), 2015 IEEE International Conference on}.\hskip
  1em plus 0.5em minus 0.4em\relax IEEE, 2015, pp. 2709--2715.

\bibitem{ma2015detect}
J.~Ma, W.~Gao, Z.~Wei, Y.~Lu, and K.-F. Wong, ``Detect rumors using time series
  of social context information on microblogging websites,'' in
  \emph{Proceedings of the 24th ACM International on Conference on Information
  and Knowledge Management}.\hskip 1em plus 0.5em minus 0.4em\relax ACM, 2015,
  pp. 1751--1754.

\bibitem{giasemidis2016determining}
G.~Giasemidis, C.~Singleton, I.~Agrafiotis, J.~R. Nurse, A.~Pilgrim, C.~Willis,
  and D.~V. Greetham, ``Determining the veracity of rumours on twitter,'' in
  \emph{International Conference on Social Informatics}.\hskip 1em plus 0.5em
  minus 0.4em\relax Springer, 2016, pp. 185--205.

\bibitem{kwon2017rumor}
S.~Kwon, M.~Cha, and K.~Jung, ``Rumor detection over varying time windows,''
  \emph{PloS one}, vol.~12, no.~1, p. e0168344, 2017.

\bibitem{borgwardt2005protein}
K.~M. Borgwardt, C.~S. Ong, S.~Sch{\"o}nauer, S.~Vishwanathan, A.~J. Smola, and
  H.-P. Kriegel, ``Protein function prediction via graph kernels,''
  \emph{Bioinformatics}, vol.~21, no. suppl\_1, pp. i47--i56, 2005.

\bibitem{buhmann2003radial}
M.~D. Buhmann, \emph{Radial basis functions: theory and implementations}.\hskip
  1em plus 0.5em minus 0.4em\relax Cambridge university press, 2003, vol.~12.

\bibitem{chen2016behavior}
W.~Chen, C.~K. Yeo, C.~T. Lau, and B.~S. Lee, ``Behavior deviation: An anomaly
  detection view of rumor preemption,'' in \emph{Information Technology,
  Electronics and Mobile Communication Conference (IEMCON), 2016 IEEE 7th
  Annual}.\hskip 1em plus 0.5em minus 0.4em\relax IEEE, 2016, pp. 1--7.

\bibitem{vydiswaran2011content}
V.~Vydiswaran, C.~Zhai, and D.~Roth, ``Content-driven trust propagation
  framework,'' in \emph{Proceedings of the 17th ACM SIGKDD international
  conference on Knowledge discovery and data mining}.\hskip 1em plus 0.5em
  minus 0.4em\relax ACM, 2011, pp. 974--982.

\bibitem{zhu2002learning}
X.~Zhu and Z.~Ghahramani, ``Learning from labeled and unlabeled data with label
  propagation,'' 2002.

\bibitem{zhu2003semi}
X.~Zhu, Z.~Ghahramani, and J.~D. Lafferty, ``Semi-supervised learning using
  gaussian fields and harmonic functions,'' in \emph{Proceedings of the 20th
  International conference on Machine learning (ICML-03)}, 2003, pp. 912--919.

\bibitem{zhou2004learning}
D.~Zhou, O.~Bousquet, T.~N. Lal, J.~Weston, and B.~Sch{\"o}lkopf, ``Learning
  with local and global consistency,'' in \emph{Advances in neural information
  processing systems}, 2004, pp. 321--328.

\bibitem{trabelsi2014mining}
A.~Trabelsi and O.~R. Zaiane, ``Mining contentious documents using an
  unsupervised topic model based approach,'' in \emph{Data Mining (ICDM), 2014
  IEEE International Conference on}.\hskip 1em plus 0.5em minus 0.4em\relax
  IEEE, 2014, pp. 550--559.

\bibitem{boididou2014challenges}
C.~Boididou, S.~Papadopoulos, Y.~Kompatsiaris, S.~Schifferes, and N.~Newman,
  ``Challenges of computational verification in social multimedia,'' in
  \emph{Proceedings of the 23rd International Conference on World Wide
  Web}.\hskip 1em plus 0.5em minus 0.4em\relax ACM, 2014, pp. 743--748.

\bibitem{derczynski2017semeval}
L.~Derczynski, K.~Bontcheva, M.~Liakata, R.~Procter, G.~W.~S. Hoi, and
  A.~Zubiaga, ``Semeval-2017 task 8: Rumoureval: Determining rumour veracity
  and support for rumours,'' \emph{arXiv preprint arXiv:1704.05972}, 2017.

\end{thebibliography}
%\newpage
\begin{IEEEbiography}[{\includegraphics[width=1in,height=1.25in,clip,keepaspectratio]{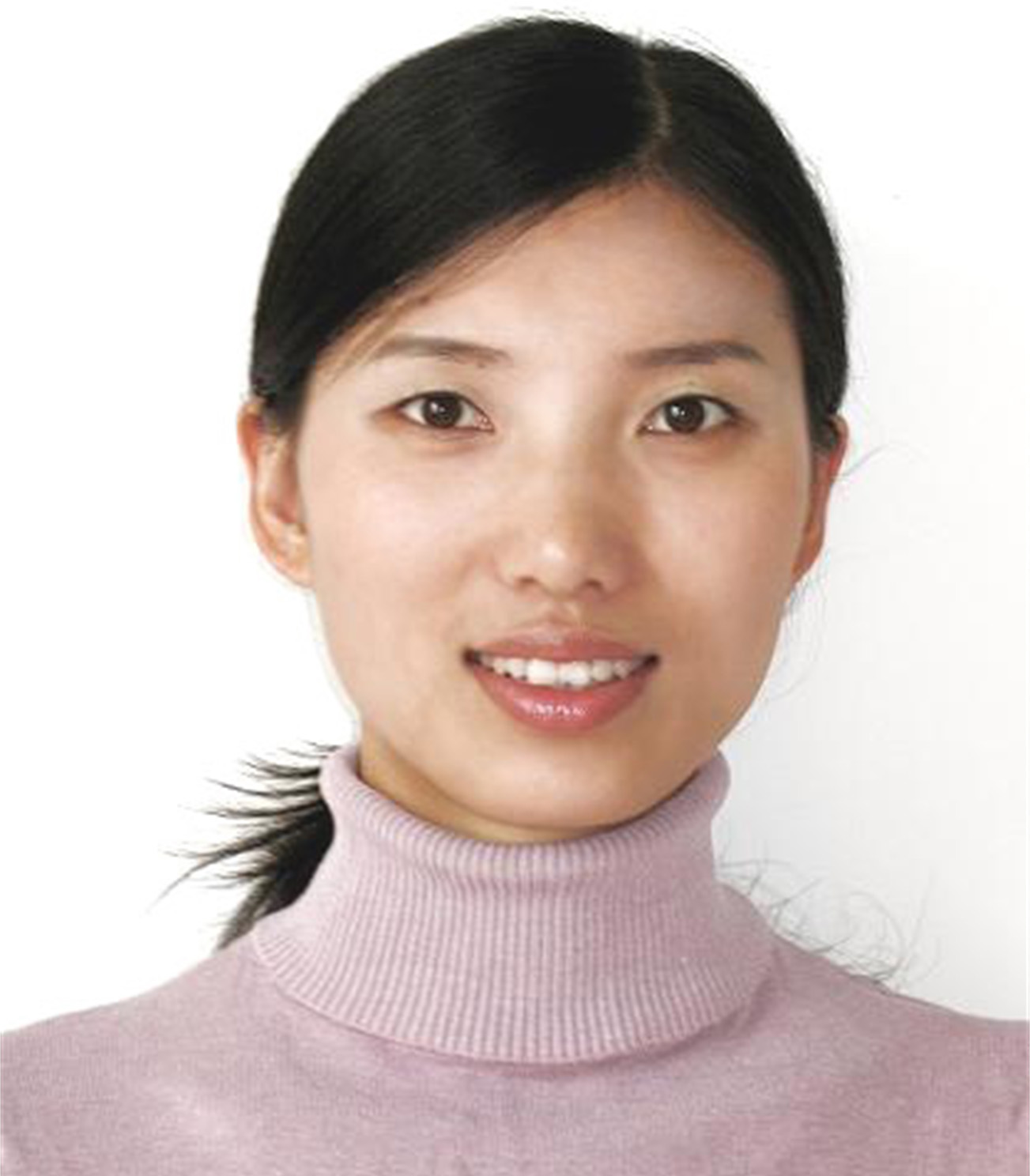}}]{Juan Cao}Institute of Computing Technology,Chinese Academy of Sciences, Center for Advanced Computing Research, Beijing, China.
University of Chinese Academy of Sciences, Beijing, China

Juan Cao received the Ph.D. degree from the Institute of Computing Technology, Chinese Academy of Sciences, Beijing, China, in 2008.
She is currently working as an Associate Professor with the Institute of Computing Technology, Chinese Academy of Sciences. Her research interests include large scale social media analysis.
\end{IEEEbiography}

\begin{IEEEbiography}[{\includegraphics[width=1in,height=1.25in,clip,keepaspectratio]{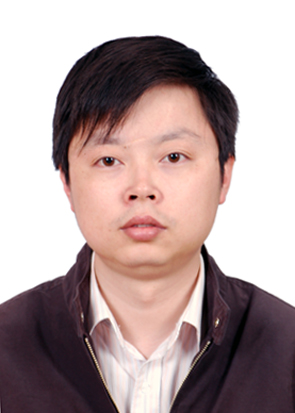}}]{Junbo Guo}Institute of Computing Technology,Chinese Academy of Sciences, Center for Advanced Computing Research, Beijing, China

Junbo Guo received the Ph.D. degree from the Institute of Computing Technology, Chinese Academy of Sciences, Beijing, China, in 2010.
He is currently working as an Senior Engineer with the Institute of Computing Technology, Chinese Academy of Sciences. His research interests include social media analysis and retrieval.
\end{IEEEbiography}

\begin{IEEEbiography}[{\includegraphics[width=1in,height=1.25in,clip,keepaspectratio]{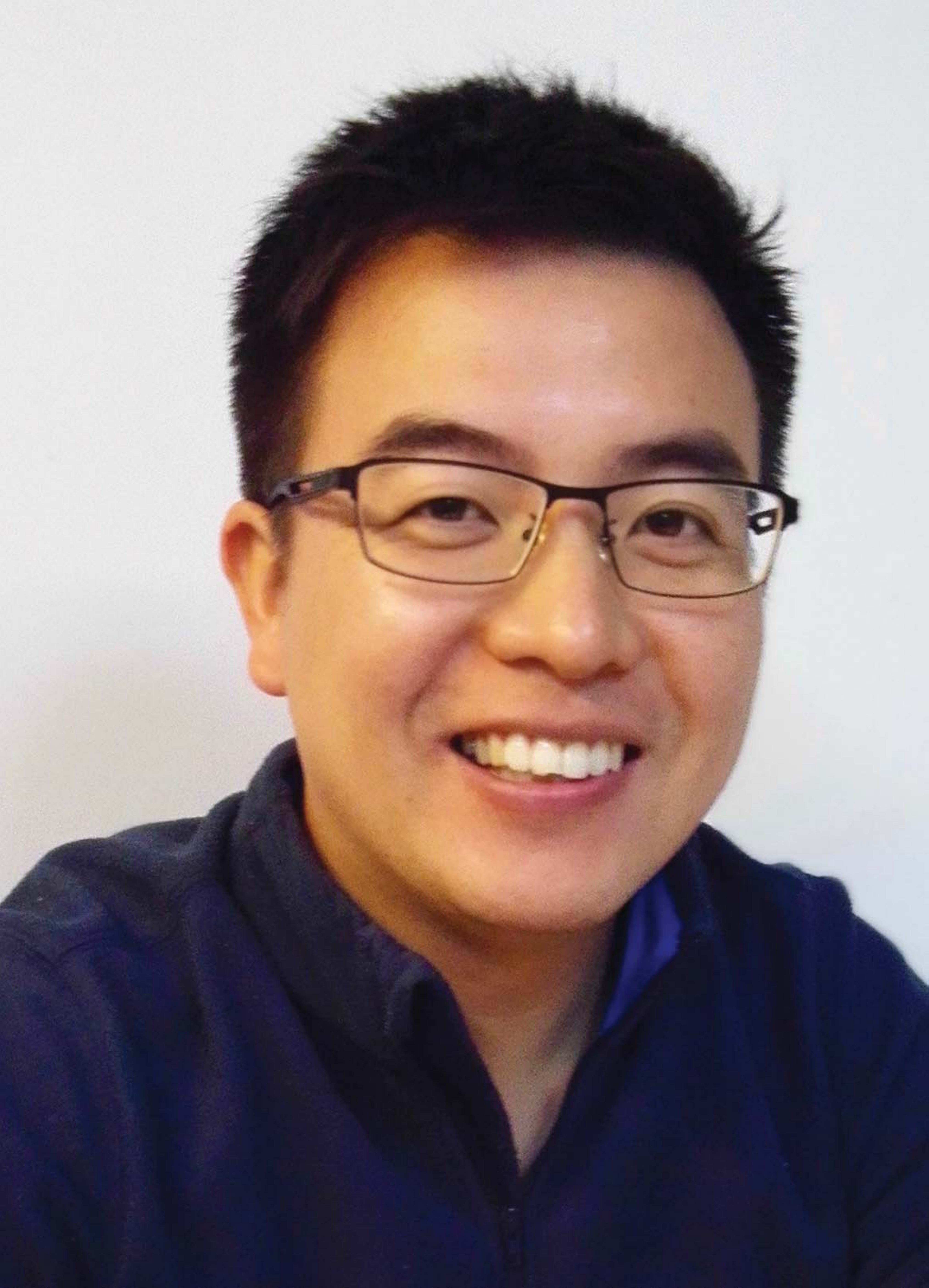}}]{Xirong Li} received the B.S. and M.E. degrees from Tsinghua University, Beijing, China, in 2005 and 2007, respectively, and the Ph.D. degree from the University of Amsterdam, Amsterdam, The Netherlands, in 2012, all in computer science.

He is currently an Associate Professor with the Key Lab of Data Engineering and Knowledge Engineering, Renmin University of China, Beijing, China. His research includes image and video retrieval.

Prof. Li was an Area Chair of ACMMM 2018 and ICPR 2016, Publication Co-Chair of ICMR 2015 and Publicity Co-Chair of ICMR 2013. He was the recipient of the ACM Multimedia 2016 Grand Challenge Award, the ACM SIGMM Best Ph.D. Thesis Award 2013, the IEEE TRANSACTIONS ON MULTIMEDIA Prize Paper Award 2012, the Best Paper Award of the ACM CIVR 2010, the Best Paper Runner-Up of PCM 2016 and PCM 2014 Outstanding Reviewer Award.

\end{IEEEbiography}

\begin{IEEEbiography}[{\includegraphics[width=1in,height=1.25in,clip,keepaspectratio]{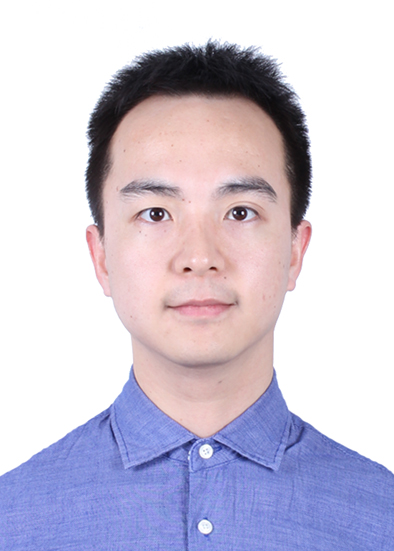}}]{Zhiwei Jin}Institute of Computing Technology,Chinese Academy of Sciences, Center for Advanced Computing Research, Beijing, China.
University of Chinese Academy of Sciences, Beijing, China

Zhiwei Jin received the B.S. degree in software engineering from Wuhan University, Wuhan, China, in 2012, and is currently working toward the Ph.D. degree at the Institute of Computing Technology, Chinese Academy of Sciences, Beijing, China, under the supervision of Prof. Y. Zhang.
His research interests include multimedia content analysis and data mining.
\end{IEEEbiography}

\begin{IEEEbiography}[{\includegraphics[width=1in,height=1.25in,clip,keepaspectratio]{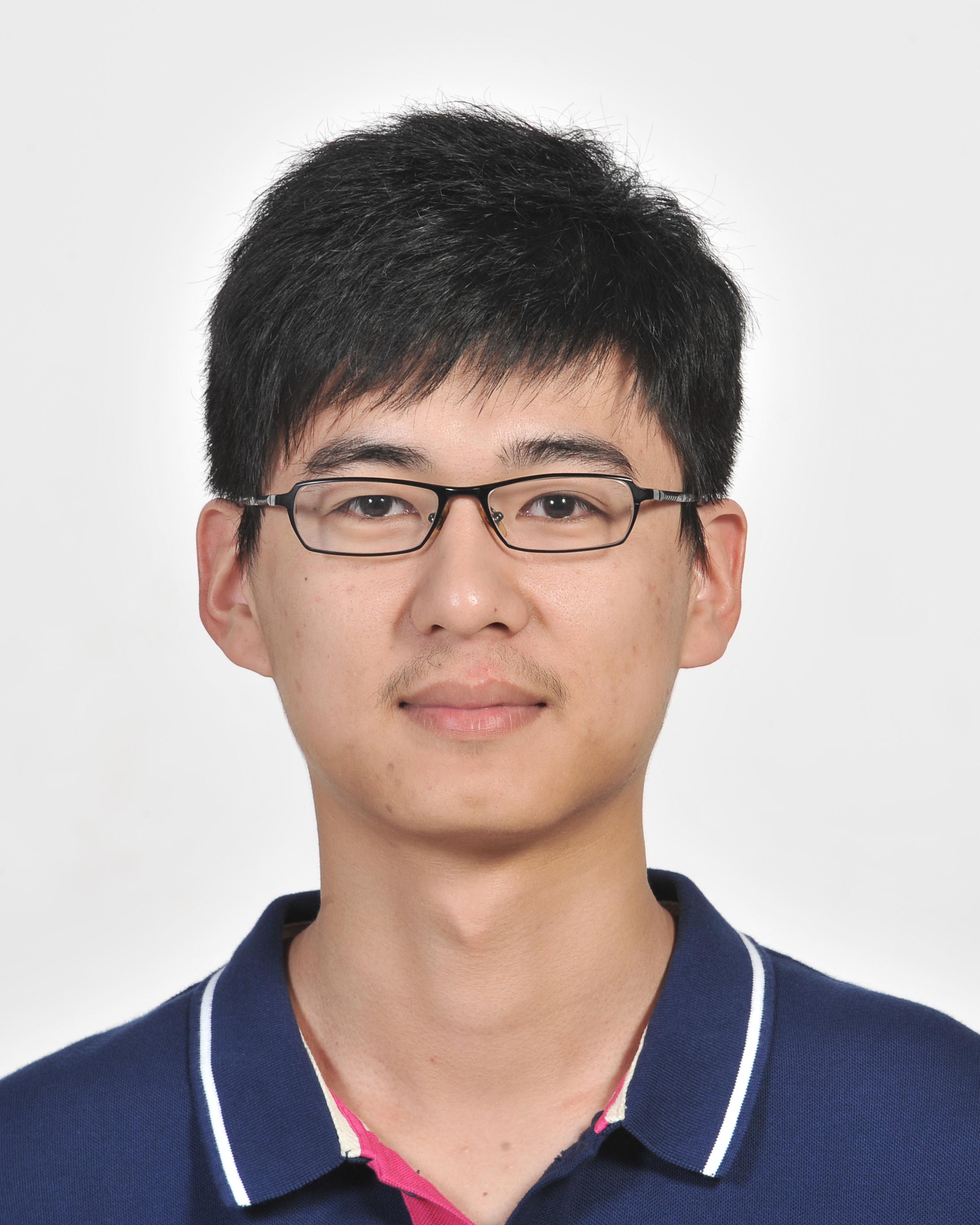}}]{Han Guo}Institute of Computing Technology,Chinese Academy of Sciences, Center for Advanced Computing Research, Beijing, China.
University of Chinese Academy of Sciences, Beijing, China

Han Guo received the B.S. degree in software engineering
from Shandong University, Jinan, China, in 2015, and is currently working toward the Master degree at the Institute of Computing Technology, Chinese Academy of Sciences, Beijing, China, under the supervision of Prof. J. Li. His research interests include multimedia content analysis.
\end{IEEEbiography}

\begin{IEEEbiography}[{\includegraphics[width=1in,height=1.25in,clip,keepaspectratio]{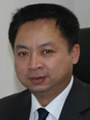}}]{Jintao Li}Institute of Computing Technology,Chinese Academy of Sciences, Center for Advanced Computing Research, Beijing, China.

Jintao Li received the Ph.D. degree from the Institute of Computing Technology, Chinese Academy of
Sciences, Beijing, China, in 1989. He is currently a Professor with the Institute of Computing Technology, Chinese Academy of Sciences. His research interests focus on multimedia technology, virtual reality technology, and pervasive computing
\end{IEEEbiography}

\end{document}